\documentclass[11pt]{article}
\pdfoutput=1
 \usepackage{mciteplus}
 \usepackage{tikz}
 \usepackage{color}
 \definecolor{darkblue}{rgb}{0.1,0.1,.7}
 \usepackage[colorlinks, linkcolor=darkblue, citecolor=darkblue, urlcolor=darkblue, linktocpage,hyperfootnotes=false]{hyperref} 
\usepackage{epsfig}
\usepackage{graphicx}
\usepackage{cite}
\usepackage{amsfonts}
\usepackage{amssymb}
\usepackage{bm}
\usepackage{latexsym}
\usepackage{amsmath}
\numberwithin{equation}{section}
\def\bq{\begin{quote}}
\def\eq{\end{quote}}
 at 10truept

\newcommand{\calo}{{\cal O}}
\newcommand{\calh}{{\cal H}}

\newcommand{\otp}{\frac{1}{2\pi}}
\newcommand{\beq}{\begin{equation}}
\newcommand{\eeq}{\end{equation}}
\newcommand{\beqa}{\begin{eqnarray}}
\newcommand{\eeqa}{\end{eqnarray}}
\newcommand{\bea}{\begin{eqnarray}}
\newcommand{\eea}{\end{eqnarray}}

\newcommand{\hf}{\frac{1}{2}}

\def\lesssim{~\mbox{\raisebox{-.6ex}{$\stackrel{<}{\sim}$}}~}
\def\roughly#1{\raise.3ex\hbox{$#1$\kern-.75em\lower1ex\hbox{$\sim$}}}

\begin{document}

\thispagestyle{empty}
\begin{titlepage}
 \setcounter{page}{0}
  \bigskip

  \bigskip\bigskip

  \bigskip

\begin{center}
{\Large \bf {Schr\"odinger evolution of two-dimensional black holes }}
    \bigskip
\bigskip
\end{center}

  \begin{center}

 \rm {Steven B. Giddings\footnote{\texttt{giddings@ucsb.edu}} }
  \bigskip \rm
\bigskip

{Department of Physics, University of California, Santa Barbara, CA 93106, USA}  \\
\rm

  \bigskip \rm
\bigskip
 
\rm

\bigskip
\bigskip

  \end{center}

\vspace{3cm}
  \begin{abstract}
This paper systematically treats the evolving quantum state for two-dimensional black holes, with particular focus on the CGHS model, but also elucidating features generalizing to higher dimensions.  This is done in Schr\"odinger picture(s), to exhibit the dynamic evolution of the state at intermediate times.  After a review of classical solutions, also connecting to descriptions of higher-dimensional black holes, it overviews the canonical quantum treatment of the full evolution, including gravitational dynamics.  Derived in an approximation to this, following  conversion to ``perturbation picture," is the  evolution of the quantum matter on the background geometry.  Features of the evolving matter state are described, based on choice of a time slicing to put the evolution into ADM form.  The choices of slicing as well as coordinates on the slices result in different quantum ``pictures" for treating the evolution.   If such a description is based on smooth trans-horizon slices, that avoids explicit reference to ultra-planckian modes familiar from traditional treatments, and exhibits the Hawking excitations as emerging from a ``quantum atmosphere" with thickness comparable to the inverse temperature.  Detailed study of the state exhibits the entanglement structure between Hawking quanta and the partner excitations inside the black hole, and the corresponding ``missing information."  This explicit description also allows direct study of the evolution and features, {\it e.g.} as seen by infalling observers,  of these partner excitations, helping to address various puzzles with them.  Explicit treatment of the evolving state, and its extension to higher dimensions, provides further connections to information theory and a starting point for study of corrections that can unitarize evolution, arising from new quantum gravity effects -- whether wormholes or something entirely different.

 \medskip
  \noindent
  \end{abstract}
\bigskip \bigskip \bigskip 

  \end{titlepage}

\tableofcontents

\newpage

\section{Introduction}

Black holes appear to be the most mysterious objects in the cosmos, and a primary reason for this is the predicted phenomenon of Hawking evaporation\cite{Hawk} and the associated question of the fate of information that they capture.

A useful way to study the information content of a black hole (BH) is through the entanglement of its quantum states with outside degrees of freedom.  A BH can build up this entanglement either by absorbing entangled matter, or it naturally builds up such entanglement via the Hawking process.  Transfer of this entanglement back to the BH environment is forbidden in a description based on local quantum field theory (LQFT), by its property of locality.  So, as is well known, if the BH disappears at the end of evaporation, this entanglement will be lost, contradicting the unitarity of quantum mechanics; the violation can be parameterized by the increase in the microscopic von Neumann entropy.  The alternative of a microscopic BH remnant remaining, preserving the entanglement, leads to even worse trouble\cite{Pres,WABHIP,Susstrouble}.  So, in short, BH decay appears to present a conflict among the most basic physical principles, which underpin LQFT.

Search for the correct resolution to this conflict has lead to significant synergies with more general studies of quantum information.  Key questions in the BH problem, and more generally, are those of the localization and flow of quantum information, and of the quantum amplitudes describing its evolution.

Since understanding localization and flow of information is key in this connection, and likely in understanding the ultimate resolution to the problem, a useful starting point is to understand the original Hawking evolution from this perspective.  While the outlines of such a story are well-understood, the original Hawking calculation\cite{Hawk}, and its many rederivations, has more of an S-matrix character, calculating amplitudes for asymptotic states rather than the real-time evolution of the process.

For a more complete understanding and connection with quantum information concepts, it seems desirable to have a description of the dynamical, real-time, evolution of the quantum state.  Within the LQFT framework we don't expect any big surprises here, and so this evolution is expected to still lead to ultimate unitarity violation.  But this also provides a foundation for addressing the central question: where and how does the true unitary quantum evolution depart from the LQFT description?  We do expect that a description of consistent unitary evolution can also be studied from the S-matrix perspective, {\it e.g.} by tracking amplitudes for and arrival times of excitations at infinity, and investigating ultimate unitarity of these amplitudes.  But, for deeper insight, it seems important to ``open the box" and understand the evolution, whatever form it takes, at a more microscopic and local level.

One purpose of this paper is to further develop such a real-time description of the LQFT evolution originally discussed  by Hawking.  This is most naturally done by working in a Schr\"odinger picture, to describe the evolving quantum state.  The basic features of such evolution are also most easily exhibited in the context of evolution of a two-dimensional BH, such as those of dilaton gravity models, though clear connections to higher-dimensional behavior can also be seen.\footnote{Earlier investigation in this direction appeared in \cite{NVU,SEHS}.  Explicit treatment of additional aspects of higher-dimensional evolution will be presented separately\cite{GiPe}. Other discussion of such evolution includes \cite{Jaco93,MeWe1,MeWe2,Jaco12,BHS,HoSi}.}

Describing the real-time evolution of amplitudes can also help give clearer answer to various questions that have arisen in the original S-matrix derivation.  One is the possible role of transplanckian effects, arising from a picture of tracing excitations back to the horizon in Hawking's original derivation.  Another is the question of how to generalize Hawking's derivation to the case of interacting theories, where such a trace-back procedure doesn't work.  One more is the question of the role and physical behavior of internal ``Hawking partner" excitations, which in a certain sense have negative energy.  

Addressing the first question, one can see explicitly that the transplanckian features of Hawking's original derivation arise from basis dependence that is an artifact of his derivation.  It is important to understand this, since Hawking's original description has suggested to some that Hawking radiation originates in a microscopic region surrounding the horizon.  A contrary viewpoint, in which it originates in a thicker ``quantum atmosphere," has been advocated in \cite{SGBoltz} (see \cite{Unru,Full,Bard} for earlier related arguments), and is supported by the real-time evolution described in a more regular basis.  

In short, it seems desirable to improve our standard of
 understanding of Hawking evolution; this also provides a more precise starting point from which to ask the question where true physical evolution departs from his description, and how.  One approach to this question is to parameterize unitary evolution as a departure from Hawking's evolution, assuming for example that it arises from corrections that are in a certain sense small.\footnote{For a brief review, see \cite{BHQU}.}

Indeed, features of the original Hawking derivation may be important clues in such a parameterization.  For example, the assumption that Hawking radiation ``is produced" in a microscopic near-horizon region suggests that its modifications also appear in this region, and that in turn predicts that infalling observers see high-energy excitations there\cite{SGTrieste,Brau,AMPS}, called a ``firewall" in  \cite{AMPS}.  A more complete picture of Hawking evolution suggests this is oversimplified.  If, on the other hand, the Hawking radiation emerges from a thicker quantum atmosphere, that is one motivation for suggesting that its modifications are operative in that broader region, as in \cite{NLvC,SGmodels,NVNL}.

Providing such a foundation for parameterization of new effects could be helpful, whatever these new effects are; candidates include modifications suggested by 't Hooft\cite{tHooft}, wormholes\cite{SSS,PSSY,AHMST,MaMa}, or other intrinsically quantum gravity effects {\it e.g.} resulting from inexactness of a spacetime description\cite{QFG}.  Various such effects might be parameterized in an effective description, such as given in \cite{NVUEFT,BHQU}, in terms of modifications to the quantum hamiltonian.  For example, it may be possible to give an effective description of wormholes this way, similarly to \cite{Cole,GiSt}. 

Another key question regards the possible observability of such new effects.  The BH conundrum appears to indicate that LQFT must be modified on the scale of the event horizon radius $R$.  If these modifications extend over a region of thickness $\calo(R)$, as suggested both by the preceding argument and by naturalness considerations, then they can extend into a region where observational signals  -- either electromagnetic or gravitational wave -- are being produced.    This suggests search for signatures of such modifications\cite{SGObs,GiPs,SGAstro,GKT}\cite{BHQU}, as an observational guide to their physics.

In outline, this paper begins with a description of 2d dilaton gravity theories, particularly that of \cite{CGHS}, and reviews classical dynamic BH solutions of the latter, also improving connections to more familiar higher-dimensional descriptions of BHs.  Section three then overviews canonical quantization of the full theory, including gravitational dynamics, to yield the evolving quantum wavefunction.  Section four describes how a perturbative approximation to this then gives evolution on a classical background; this involves a change of quantum picture, analogous to that to interaction picture, but here to what is called ``perturbation picture."

Section five begins the in-depth treatment of the evolving quantum state of matter on the classical BH background.  It first overviews different BH coordinatizations, as well as slicings needed to give an ADM description of the dynamics.  This is followed by treatment of the evolution of the state.  This is done in a Schr\"odinger picture; different slicings and coordinatizations in fact yield what amount to different such {\it pictures}.  In a regular description, based on a smooth trans-horizon slicing, one can find the explicit form of the hamiltonian, and see both the smooth structure of the quantum atmosphere as well as the connection to the more singular energy eigenstate basis.  This leads to an explicit treatment of the Hawking evolution which supports many of the features which have been generally appreciated, but in a more complete description, which is also readily generalized to the interacting case.  In particular, comparison of different descriptions clearly exhibits the development of entanglement between Hawking radiation and BH states.  The description of these states is connected to the properties of internal Hawking partners, which are explicitly studied in section six, for example   from the viewpoint of an infalling observer.  This provides helpful perspective on puzzles involving these excitations.  Section seven concludes, and also discusses connections with further directions, such as parameterizing unitarizing deviations from Hawking evolution. 

Appendices give a brief overview of useful ADM expressions, as well as an introduction to ``perturbation picture."

Sections three and four can be skipped by those readers only interested in the description of evolution of the quantum state on a fixed BH geometry.

\section{Classical theories and black hole solutions}

\subsection{Dilaton gravity}

Two-dimensional dilaton gravity provides a model for investigating many (though not necessarily all) aspects of black hole evolution.  One motivation for it is to consider the radially-symmetric truncation of four-dimensional Einstein gravity, assuming a metric of the form
\beq
ds^2=g_{\mu\nu}(t,r)dx^\mu dx^\nu + \frac{1}{\lambda^2} e^{-2\phi(t,r)} d\Omega_2^2\ ;
\eeq
here $x^\mu = (t,r)$,  $ d\Omega_2^2$ is the line element on $S^2$, and $\lambda$ is a constant with dimension of inverse length.  
The Einstein action then reduces to
\beq
S_E=\frac{1}{4G\lambda^2} \int d^2x \sqrt{|g|} e^{-2\phi} \left[R+ 2(\nabla \phi)^2 + 2\lambda^2 e^{2\phi}\right]\ .
\eeq
A more general class of dilaton gravity theories has action
\beq
S=-\otp\int d^2x \sqrt{|g|}\left[K(\phi) (\nabla\phi)^2 + \frac{1}{2} \Phi(\phi) R + U(\phi)\right]
\eeq
and has been investigated beginning in the 1990s (see \cite{QTDG} and references therein).

A particularly tractable model is motivated by the 2d reduction of the string action, together with one (or more) minimally coupled scalar fields\cite{CGHS},
\beq\label{CGHSact}
S=\otp \int d^2 x\sqrt{|g|} e^{-2\phi}\left[R+4(\nabla\phi)^2 + 4\lambda^2\right] - \frac{1}{4\pi}\int d^2 x\sqrt{|g|} (\nabla F)^2 + S_\partial\ ,
\eeq
where $\lambda$ is again a constant with dimensions of inverse length.\footnote{Here we use the traditional normalization\cite{CGHS} of the action, which matches certain conventions in study of 2d  theories; this results in  factors of $2\pi$ relative to standard 4d conventions.  For example, a canonically normalized scalar field, such as $\phi$ in appendix 
\ref{ADMA}, is related as $F=\sqrt{2\pi} \phi$.}
The surface term in \eqref{CGHSact} can be fixed by the requirement\cite{GiHa} that the action only depend on first derivatives of the metric, ensuring a good variational principle with vanishing metric perturbation on the boundary.  This gives\footnote{For recent discussion of other boundary conditions in the context of Jackiw-Teitelboim gravity, see \cite{GIKY}.} 
\beq
S_\partial = \frac{1}{\pi} \int_\partial dl \, e^{-2\phi} K_\partial\ ,
\eeq
where $dl$ is the boundary length element, and $K_\partial$ the extrinsic curvature of the boundary.  

The action \eqref{CGHSact}, commonly called the CGHS model, has a general class of exact solutions describing 2d black hole formation, and their evaporation can also be simply treated.

\subsection{General classical solution}

The equations of motion arising from varying the metric and dilaton in \eqref{CGHSact} are
\beq\label{meteqn}
2\left\{\nabla_\mu\nabla_\nu \phi - g_{\mu\nu}\left[\square\phi -(\nabla\phi)^2+\lambda^2\right]\right\} = e^{2\phi}T_{\mu\nu}
\eeq
and
\beq
\square\phi   -(\nabla\phi)^2+\lambda^2+ \frac{R}{4} =0\ ,
\eeq
where
\beq
T_{\mu\nu}=\frac{1}{2}\left[\nabla_\mu F\nabla_\nu F - \frac{1}{2} g_{\mu\nu} (\nabla F)^2\right]
\eeq
is the matter stress tensor.\footnote{The traditional normalization used here differs from the usual canonical definition, $T^{\rm Canon}_{\mu\nu}=-(2/\sqrt{|g|})\delta S/\delta g^{\mu\nu}$.  They are related by $T_{\mu\nu}^{\rm Canon} = T_{\mu\nu}^{\rm CGHS}/\pi$.}  One sees that $e^{2\phi}$ is an effective coupling analogous to Newton's constant.  These have a vacuum solution
\beq\label{LDV}
g_{\mu\nu} = \eta_{\mu\nu}\quad ,\quad \phi= -\lambda r\ ,
\eeq
with spatial coordinate $r$, known as the linear dilaton vacuum.  

More general solutions are most easily analyzed by using a diffeomorphism to put the metric in conformal gauge,
\beq
ds^2= -e^{2\rho} dx^+ dx^-\ ,
\eeq
with $x^\pm = t\pm x$.  Using the relation $\sqrt{|g|} R = -2\square\rho$, the gravitational action becomes
\beq
S_{g\phi} = \frac{1}{2\pi} \int d^2x \left[2\nabla(\rho-\phi)\cdot\nabla e^{-2\phi} + 4\lambda^2 e^{2(\rho-\phi)}\right]\ .
\eeq
Varying $e^{-2\phi}$ shows that $\rho-\phi$ behaves as a free field, with general solution\footnote{Formally, this simplification arises from the symmetry $e^{-2\phi}\rightarrow e^{-2\phi} + \epsilon$, $\rho-\phi\rightarrow \rho-\phi$.  The action also has classical conformal symmetry.}
\beq
\rho-\phi= \frac{1}{2} \left[w_+(x^+) + w_-(x^-)\right]\ .
\eeq
Clearly $w_\pm$ are shifted by a conformal diffeomorphism, which changes the gauge.
Varying $\rho-\phi$ then gives an equation for $\phi$, with general solution
\beq
e^{-2\phi} = u_+(x^+) + u_-(x^-) - \lambda^2 \int^{x^+} dx^+ e^{w_+} \int^{x^-} dx^- e^{w_-}\ .
\eeq

The remaining $++$ and $--$ equations of \eqref{meteqn} determine $u_\pm$ in terms of the matter distribution.
They take the form 
\bea
-(4\partial_+\rho\partial_+\phi -2\partial_+^2\phi) &=& e^{2\phi}\, T_{++}\cr
-(4\partial_-\rho\partial_-\phi -2\partial_-^2\phi) &=& e^{2\phi}\, T_{--}
\eea
and have solution
\beq\label{ueq}
u_{\pm}= \frac{M_\pm}{\lambda} -  \int dx^\pm e^{w_\pm} \int dx^\pm e^{-w_\pm} T_{\pm\pm}
\eeq
where $M_\pm$ are  integration constants.  A useful simplification is to work in units with $\lambda=1$.
With vanishing $T_{\mu\nu}$ and $M_\pm=0$, we have $u_\pm=0$ and the choice of gauge $w_+=x^+$, $w_-=-x^-$ reproduces the linear dilaton vacuum \eqref{LDV}.  

\subsection{Black hole spacetimes}\label{BHS}

A black hole is formed by a left moving  wave
$F=F_0(x^+)$ incident on the linear dilaton vacuum.  Specifically, with the preceding choice of gauge, \eqref{ueq} then gives
\beq
u_+ = M(x^+) - e^{x^+} \Delta(x^+)\quad ,\quad
\eeq
where 
\beq
M(x^+) = \int_{-\infty}^{x^+} dx^+ T_{++}\quad , \quad \Delta(x^+) =  \int_{-\infty}^{x^+} dx^+ e^{-x^+} T_{++}\ ,
\eeq
resulting in dilaton
\beq\label{dilsoln}
e^{-2\phi} = M(x^+) + e^{x^+} \left[e^{-x^-} - \Delta(x^+)\right]
\eeq
and metric
\beq\label{BHsoln}
ds^2 = - \frac{dx^+ dx^-}{1+ M(x^+) e^{x^--x^+} - \Delta(x^+) e^{x^-}}\ .
\eeq
The special case of a delta-function shock wave 
$
T_{++}=M\delta(x^+-x_0^+)
$
gives $\Delta = M e^{-x_0^+}$.

\begin{figure}[!hbtp] \begin{center}
\includegraphics[width=10cm]{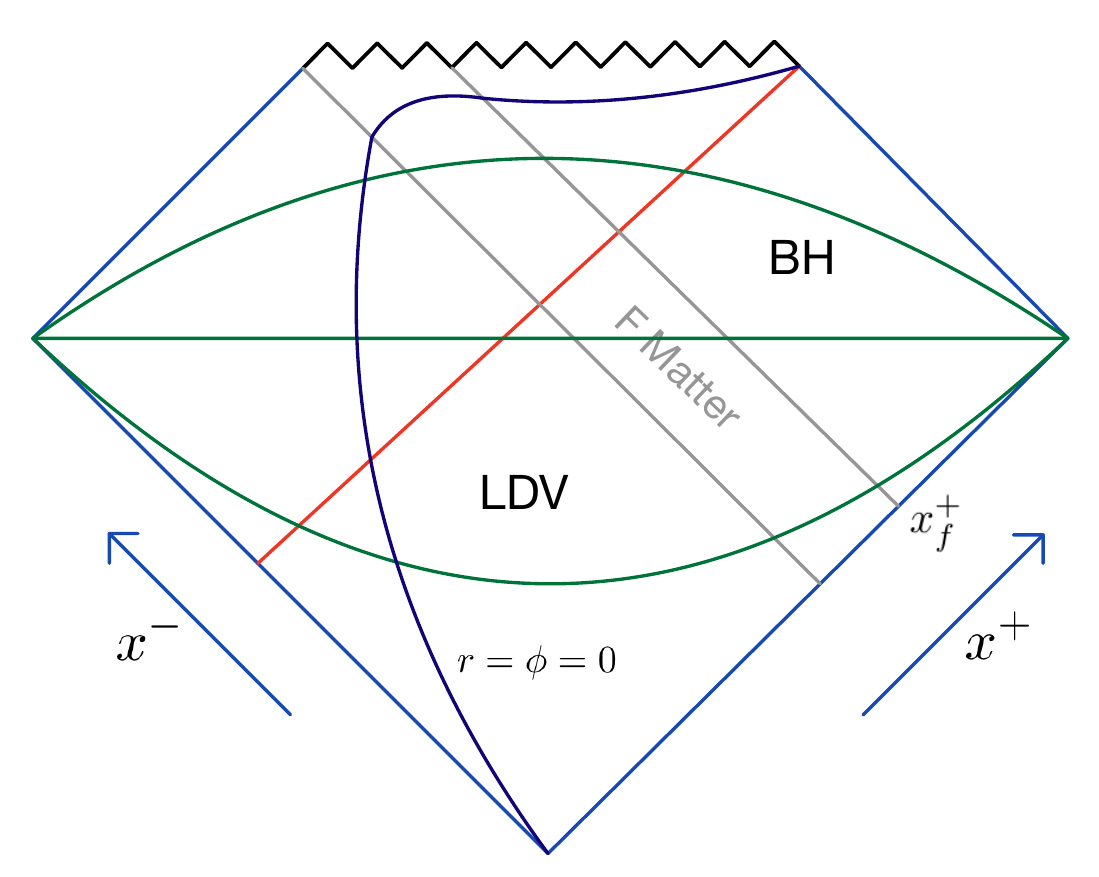}
\end{center}
\caption{A Penrose diagram for the black hole \eqref{BHsoln} formed from $F$ matter incident on the linear dilaton vacuum.  The incident  field turns off at advanced time $x_f^+$, and  the subsequent solution is a vacuum black hole.  Also shown are the horizon in red, a
set of spatial slices in green, and the line $r=\phi=0$, which can be thought of as an approximate boundary for the strong-coupling region.}
\label{FigPen}
\end{figure} 

A Penrose diagram for the solution \eqref{BHsoln} is shown in Fig.~\ref{FigPen}.
To understand the interpretation of the metric \eqref{BHsoln} as a black hole, one  can choose a gauge preserving the spatial coordinate definition $r=-\phi$.  Then a coordinate transformation to the 
coordinates $(r,x^+)$ yields the metric
\beq\label{Vaidform}
ds^2= -\left[1-M(x^+) e^{-2r}\right] dx^{+2} + 2 dx^+ dr\ .
\eeq
which is a black hole metric in Vaidya form\cite{Vaid}. Define
\beq\label{fdef}
f(x^+,r)= 1-M(x^+)e^{-2r}\ 
\eeq
and suppose that $T_{\mu\nu}$ vanishes after advanced time $x_f^+$, so that $M(x^+)=M$ is subsequently constant.   
Then, if for $x^+>x^+_f$ we define the  ``tortoise coordinate"
\beq
r_*=\int \frac{dr}{f(r)} =\hf \ln\left\vert e^{2r} - M\right\vert \ ,
\eeq
 the substitution $x^+=t+r_*$ puts the $x^+>x_f^+$ metric in standard Schwarzschild form,
\beq\label{Schmet}
ds^2 = -f(r) dt^2 + \frac{dr^2}{f(r)}\ ,
\eeq
with horizon at $r=R=\ln M/2$. 

\begin{figure}[!hbtp] \begin{center}
\includegraphics[width=12cm]{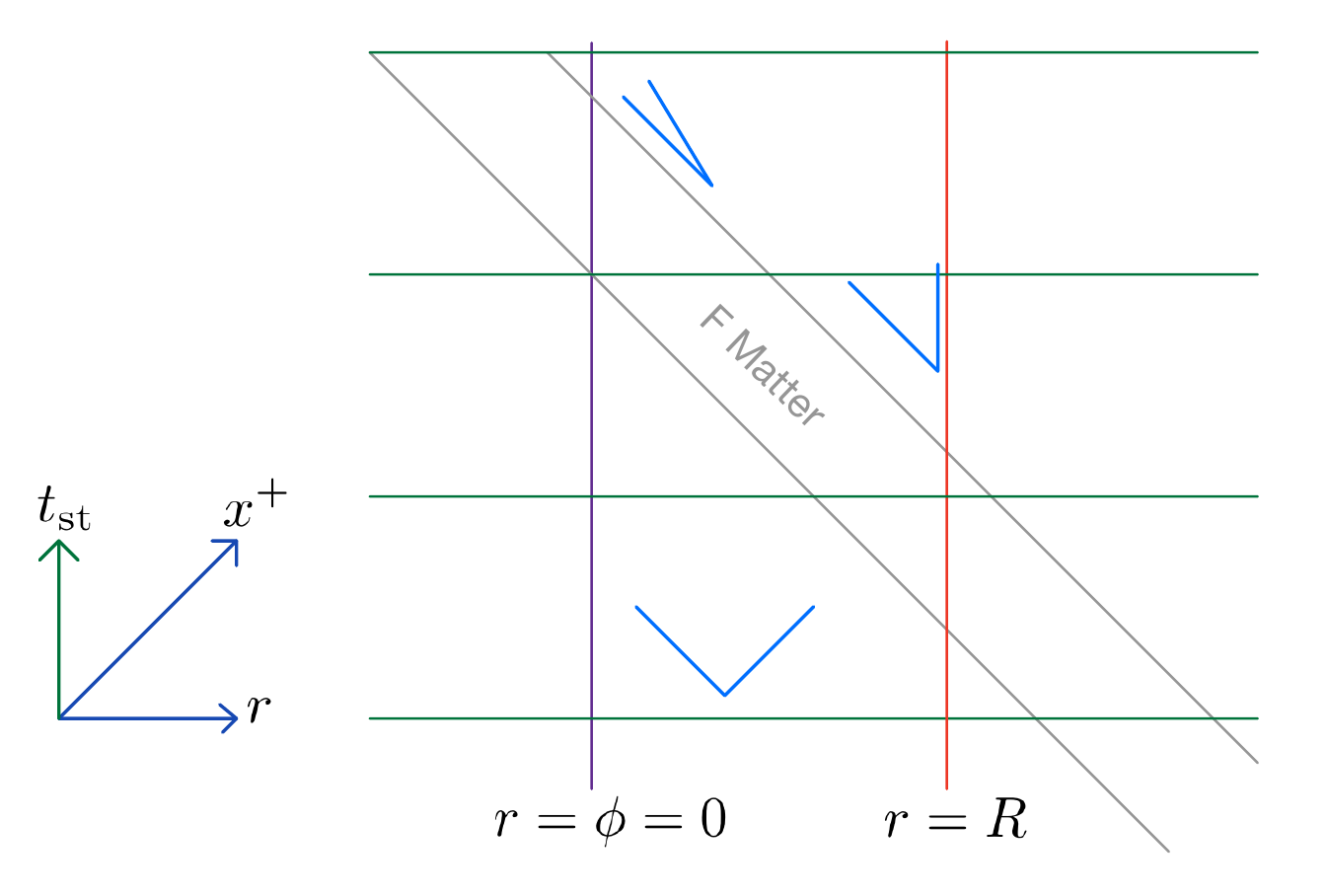}
\end{center}
\caption{An Eddington-Finkelstein diagram for the same spacetime as Fig.~\ref{FigPen}.  Spatial slices are taken to be the ``straight" slices 
described in \ref{SADM}, with time coordinate denoted $t_{\rm st}$.  The line $r=\phi=0$ and horizon at $r=R$ are shown, as are representative light cones in light blue.}
\label{FigEF}
\end{figure}

The function $f$ in \eqref{fdef}  for $x^+>x_f^+$ is the 2d version of the dimension $D>3$ function $f=1-(R/r)^{D-3}$.
The metric \eqref{BHsoln} provides a good global description of the spacetime, which has a singularity at $r=-\infty$, rather than at $r=0$ for higher $D$.   However, as $r=-\phi$ approaches zero, the effective coupling $e^{2\phi}$ becomes $\calo(1)$, and so that might correspondingly be interpreted as placing an endpoint on the semiclassical spacetime.  An Eddington-Finkelstein diagram for the spacetime is shown in Fig.~\ref{FigEF}.

\section{Full canonical evolution of the wavefunction}

A general approach to investigate the quantum evolution of a spacetime is to work with a time slicing (which ultimately is part of  a choice of gauge), and describe the evolution with respect to that slicing.  
Before turning to quantum evolution for black holes, we  next give a general description of  the evolving wavefunction in dilaton gravity.  In a general time slicing the metric takes the ADM form \cite{ADM}
\beq\label{ADMmet}
ds^2 = -N^2 dt^2 + q (dx + N^x dt)(dx+ N^x dt)\ 
\eeq
where $q=q_{xx}=g_{xx}$ is the spatial metric component.  Description of the canonical evolution begins by identifying the momenta and hamiltonian.

We first review this for the matter action, which in ADM variables is
\beq
S_F= \frac{1}{4\pi} \int dt dx \sqrt q N \left[  \frac{1}{N^2}\left(\dot F - N^x\partial_x F\right)^2 - q^{-1}(\partial_x F)^2 \right]\ ;
\eeq
this and other useful formulas in ADM variables are summarized in Appendix \ref{ADMA}.
We define the canonical momentum as
\beq\label{Fmom}
\Pi=  \frac{1}{\sqrt q}\frac{\delta S}{\delta\dot F} = \frac{1}{2\pi N} \left(\dot F - N^x\partial_x F\right)=  \frac{1}{2\pi}\, \partial_n F 
\eeq
where 
\beq\label{ndef}
n^\mu = (1,-N^x)/N\ 
\eeq
is the normal to a constant-$t$ slice. 
The canonical commutation relations are
\beq\label{CCR}
[\Pi(x,t),F(x',t)] = -\frac{i}{\sqrt q}\delta(x-x')\ .
\eeq
In canonical variables, the action then becomes
\beq\label{Fcanact}
S_F = \int dt dx\sqrt{q} \left(\Pi\dot F  -  \calh_F\right)\ ,
\eeq
with hamiltonian density
\beq\label{HF}
\calh_F =  \frac{N}{4\pi} \left[\left(2\pi \Pi\right)^2 + \frac{(\partial_x F)^2}{q}\right] +  N^x\Pi \partial_x F\ .
\eeq
The Schr\"odinger evolution of an initial state $|\psi_0\rangle$ is then of the form
\beq
|\psi,t\rangle = \exp\left\{ -i \int dt dx\sqrt q\, \calh_F\right\}|\psi_0\rangle\ .
\eeq

Generalization to include evolution of the metric and dilaton requires reexpressing the gravitational action \eqref{CGHSact} in canonical form.\footnote{For a related canonical analysis, see \cite{KRV}.}  The general expression for the Ricci scalar in ADM variables (see Appendix \ref{ADMA}) simplifies in two dimensions to a total derivative,
\beq\label{RADM}
\sqrt{|g|} R= -2\partial_x\left(\frac{\partial_x N - K N_x}{\sqrt q}\right) - 2\partial_t\left(\sqrt q K\right)
\eeq
where $K$  is the scalar extrinsic curvature of constant-$t$ slices,
\beq
K=q^{xx}K_{xx}=-\frac{1}{2qN}\left(\dot q -2D_xN_x\right)= -\frac{1}{2q} \partial_n q + \frac{\partial_xN^x}{N}\ .
\eeq
Inserting \eqref{RADM} in the action \eqref{CGHSact} and integrating by parts, the surface terms cancel $S_\partial$, and the gravidilaton action becomes
\beq\label{Act2d}
S_{g\phi} =\frac{2}{\pi}\int d^2x \sqrt{q} N e^{-2\phi} \left[  -K \partial_n \phi -  \left(\partial_n \phi \right)^2 - \frac{\partial_x N \partial_x \phi}{qN}+\frac{(\partial_x \phi)^2}{q}  + \lambda^2 \right]\ .
\eeq

The canonical momenta are then
\beq\label{Piq}
\Pi_q =\frac{1}{\sqrt q}  \frac{\delta S}{\delta\dot q} = \frac{1}{\pi q}e^{-2\phi} \partial_n \phi
\eeq
\beq\label{Pip}
\Pi_\phi = \frac{1}{\sqrt q}\frac{\delta S}{\delta\dot \phi}=-\frac{2}{\pi}e^{-2\phi} (K+2\partial_n\phi)\ .
\eeq
These together with \eqref{Act2d} then give us the canonical form of the action,
\beq\label{canact}
S_{g\phi} = \int d^2 x\sqrt q \left(\Pi_q\dot q  +\Pi_\phi\dot\phi  -\calh_{g\phi}          \right)
\eeq
with hamiltonian density
\bea
\calh_{g\phi} &=& N\left\{\pi e^{2\phi}q\Pi_q\left(\Pi_\phi+2q\Pi_q \right) +\frac{2}{\pi} e^{-2\phi}\left[\frac{1}{q}(\partial_x\phi)^2- \lambda^2 -D^2\phi \right]\right\}\cr
&&+ N^x \left[\Pi_\phi\partial_x\phi  -2\partial_x(q\Pi_q) \right]\cr
&&- D_x\left(\frac{N}{\pi q}\partial_x e^{-2\phi}-2 q\Pi_q N^x\right) 
\eea

The full hamiltonian takes the form
\beq\label{totham}
H = \int dx\sqrt q \left(\calh_{g\phi} + \calh_F\right) = \int dx \sqrt q\left( N{\cal C}_N + N^x {\cal C}_x\right) + H_\partial
\eeq
where
\beq\label{CN}
C_N= -\frac{2}{\pi}e^{-2\phi}\left(D^2\phi-\frac{1}{q}(\partial_x\phi)^2  +\lambda^2\right) + \pi e^{2\phi} q\Pi_q\left( \Pi_\phi+2q\Pi_q \right) + \frac{1}{4\pi }\left[ \left(2\pi \Pi\right)^2 +\frac{1}{q}(\partial_x F)^2  \right]
\eeq
and
\beq\label{Cx}
 C_x= \Pi_\phi\partial_x\phi-2\partial_x(q\Pi_q) + \Pi \partial_x F
\eeq
are the constraints, and
\beq
H_\partial = 
\left(-\frac{N}{\pi}\partial_x e^{-2\phi} + 2q\Pi_q N_x\right)_\infty
\eeq
is the boundary contribution to the hamiltonian.  
For the black hole solution \eqref{Schmet}, this gives the value
\beq\label{bdyH}
H_\partial=\frac{M}{\pi} - {\rm infinite\ constant}\ .
\eeq

The equations of motion for $N$ and $N^x$ give the constraint equations
\beq\label{constraints}  
C_N=C_x=0\ .
\eeq
If and only if these vanish, the hamiltonian \eqref{totham} is given by this surface term.  On the other hand, for a generic state, the hamiltonian \eqref{totham} can be written using its first expression 
as the integral over a constant-$t$ slice.  The term $\calh_F$ can be rewritten in terms of the stress tensor,\footnote{The unusual normalization in this and \eqref{bdyH} arises from  historical 2d conventions; see footnotes 4 and 6.}
\beq
\calh_F = \frac{1}{\pi}T_{nt}
\eeq
suggesting that $\calh_{g\phi}$  likewise be interpreted as due to an effective gravidilaton stress tensor.
The full state evolves as
\beq\label{stateevol}
|\psi,t\rangle = \exp\left\{ -i \int dt H \right\}|\psi_0\rangle\ ,
\eeq
and so is governed by $H_\partial$ for states annihilated by the constraints.  Of course, determining the full evolution requires gauge-fixing conditions, relating $N$ and $N^x$ and the other variables.

The classical equations of the preceding section are  equivalent to the equations derived from the canonical action \eqref{Fcanact}, \eqref{canact}.  These include  the constraints \eqref{constraints} and the equations arising from varying $\Pi_q$ and $\Pi_\phi$, which reproduce \eqref{Piq} and \eqref{Pip}.  Varying $q$ and $\phi$ gives the remaining equations in the gravidilaton sector.

The gravitational backreaction induced by matter, and specifically the gravitational dressing of matter excitations, can be found by solving the constraints.  As with \eqref{meteqn}, we see that this backreaction is small with the coupling $e^{2\phi}$.  For configurations where this parameter is small, we may work to leading perturbative order in which matter fluctuations propagate on a classical background.  We will leave more complete treatment of the quantum gravitational field and dynamics for future work,\footnote{For related treatment of JT gravity, see \cite{HaWuA}.} and largely focus on this approximation for the remainder of this paper.

\section{Perturbative treatment of evolution; perturbation picture}

For example, one may consider the quantum state corresponding to an $F$-wave incident from infinity on the state corresponding to the linear dilaton vacuum. The state can be written
\beq
|F_0\rangle = e^{-\frac{i}{\pi}\int dx F_0(x+T) \partial_+F(x,T)}|0,{\rm LDV}\rangle\ 
\eeq
in the asymptotic limit of vanishing coupling, achieved by taking time $T$ large and negative.  Here
$F_0(x^+)$ gives the profile of the wave; one can readily check that $\langle F_0|F(x)|F_0\rangle = F_0(x^+)$.  As the wave propagates into finite values of the dilaton background, the constraints \eqref{constraints} must be solved to determine the gravidilaton part of the wavefunction.

The quantum evolution may be studied as a perturbation about the classical solution that $F_0$ produces.  Specifically, let $g_0$ and $\phi_0$ be the corresponding solution,
given by \eqref{dilsoln}, \eqref{BHsoln}.  Then, the quantum variables may be expanded as $F=F_0+f$, $g=g_0+h$, and $\phi=\phi_0 + \varphi$.
If the fields and momenta are collectively denoted $\Phi_I$, $\Pi_I$, and their perturbations $\tilde \Phi_I$, $\tilde \Pi_I$, the hamiltonian has the corresponding expansion
\bea\label{Hexp}
H&=&H_0 + \tilde \Phi_I\frac{\partial H}{\partial \Phi_I}{\vert_0} + \tilde \Pi_I\frac{\partial H}{\partial \Pi_I}\vert_0 + \tilde H\cr
&=& H_0 - \tilde \Phi_I \dot\Pi_{0I} + \tilde \Pi_I \dot \Phi_{0I} + \tilde H\ ,
\eea
where $H_0$ is the hamiltonian for the classical solution, $\tilde H$ has all contributions of second and higher order in the perturbations, and Hamilton's equations have been used in the second line.  

The evolving state \eqref{stateevol}  has an overall phase $\exp\{-i H_0 t\}$, together with evolution by the perturbation hamiltonian $\tilde H$.  For time-dependent backgrounds, 
the linear terms in \eqref{Hexp} can be seen to induce a change of picture, analogous to that between Schr\"odinger and interaction pictures; this is explained further in Appendix \ref{PP}.\footnote{For previous discussion, see also \cite{MaroUH}.}  We refer to the new picture as ``perturbation picture." In the original   Schr\"odinger picture, the operators $\Phi_I$ are time independent, and the state evolves according to \eqref{stateevol}.  In perturbation picture the perturbation operators $\tilde \Phi_I$ are time independent, and the state evolves as
\beq
|\psi,t\rangle = \exp\left\{ -i \int dt \tilde H \right\}|\psi_0\rangle\ .
\eeq
An easy way to understand the need for the change of picture is the inconsistency of time-independence of $\Phi_I$ and $\tilde \Phi_I$ if $\Phi_I=\Phi_{0I}(t)+\tilde \Phi_I$.  

If we perturb about the classical collapse solution of \eqref{BHsoln}, or about the linear dilaton vacuum, the constraints \eqref{CN}, \eqref{Cx}, \eqref{constraints}, together with the gauge conditions, determine the metric and dilaton fluctuations $h,\varphi$ in terms of the matter fluctuation $f$.  One sees directly that the result is that $h$ and $\varphi$ are of order $e^{2\phi_0}$.  In turn, this means that for small effective coupling $e^{2\phi_0}$, the leading order contribution to $\tilde H$ is simply that of the matter field, namely \eqref{HF} with $F\rightarrow f$ and evaluated in the background metric.  One can investigate higher-order contributions to the evolution (which includes backreaction of the $f$ perturbations on the metric and dilaton), but for now will focus on this leading-order evolution.

\section{Leading-order evolution: matter on background}

We now turn to the leading order perturbative evolution.  We have argued that this evolution is governed by an ADM matter hamiltonian \eqref{HF}, defined given a slicing of the background spacetime of the classical BH solution,  \eqref{BHsoln}.  Some initial investigation of this evolution appeared in \cite{SEHS}, and the present discussion will expand on that, further elucidating some of its features. 

\subsection{Classical black hole background}

We begin by recapping and extending the description of the classical BH background, and then discussing foliations of it that can be used to describe hamiltonian evolution.

\subsubsection{Coordinate descriptions}
\label{Params}

The  collapsing BH solution was given by \eqref{BHsoln} in $(x^+,x^-)$ coordinates, or   by \eqref{Vaidform} in  Vaidya coordinates $(x^+,r)$.  
Notice that  the dilaton solution \eqref{dilsoln} and the coordinate choice $r=-\phi$  implies the coordinate relation
\beq\label{basicc}
e^{2r} = M(x^+) + e^{x^+} \left[e^{-x^-} - \Delta(x^+)\right]\ .
\eeq
Both of these coordinate descriptions are smooth across the horizon. In the coordinates $x^\pm$, the event horizon corresponds to
\beq
e^{-x^-}=\Delta\ ,
\eeq
where $\Delta = \Delta(x_f^+)$, and the singularity is at $r=-\infty$, or infinite $\phi$, corresponding to
\beq
e^{-x^-_s}=\Delta(x^+) - M(x^+) e^{-x^+}\ .
\eeq
In the post-formation vacuum region $x^+>x^+_f$, the horizon was found to be alternately described as lying at  $r=R=\ln M/2$.

Other coordinate descriptions are also useful.  For $x^+>x^+_f$ the Schwarzschild form of the metric \eqref{Schmet} was found with time defined  by $x^+=t+r_*(r)$.  The corresponding chiral coordinate
$y^-=t-r_*(r)$ then describes the region outside the horizon.  This exterior coordinate may equivalently be defined via
\beq\label{ydef}
e^{-y^-}= e^{-x^-}-\Delta,
\eeq
and this definition can be extended to $x^+<x^+_f$ as well.  
An analogous expression
\beq\label{yhatdef}
e^{\hat y^-}= \Delta -e^{-x^-}
\eeq
describes  the interior.  In the vacuum region $x^+>x_f^+$ the metric then takes the form
\beq
ds^2=-dx^+dy^-(1-Me^{-2r})= -dx^+ d\hat y^- (Me^{-2r}-1)\ 
\eeq
which is  asymptotically flat as $r\rightarrow\infty$.
The coordinates $y^-$ and $\hat y^-$ are singular at the horizon.  

Another choice of smooth coordinates at the horizon is that of Kruskal,
\beq\label{Krusk}
X^+=e^{x^+}\quad ,\quad X^-=\Delta - e^{-x^-}=-e^{-y^-}\ ,
\eeq
in terms of which the final BH metric becomes
\beq
ds^2=-\frac{d X^- dX^+}{M-X^+X^-}\ .
\eeq

\subsubsection{Slices and ADM parameterizations}
\label{SADM}

The description of dynamical evolution in this BH spacetime depends on a choice of time slices.  In general, these can be parameterized as 
\beq
x^\mu={\cal X}^\mu(t,x)
\eeq
where $t$ labels the slices and $x$ is a spatial coordinate along the slices.  Working in the background of the BH \eqref{BHsoln} and eliminating $x$, one can alternately write
\beq
x^+={\cal S}(t,x^-)
\eeq
to specify the general slicing.  Once the BH has formed, at $x^+> x_f^+$, the solution \eqref{Vaidform} has translation symmetry under shifts of $x^+$.  We expect the evolution to simplify if we choose slices respecting this symmetry, and these are determined by\cite{NVU}\cite{SEHS}
\beq\label{slicedef}
x^+= t+ S(r)
\eeq
for a given ``slice function" $S(r)$.  We refer to this type of slicing as a {\it stationary slicing.}

\begin{figure}[!hbtp] \begin{center}
\includegraphics[width=9.25cm]{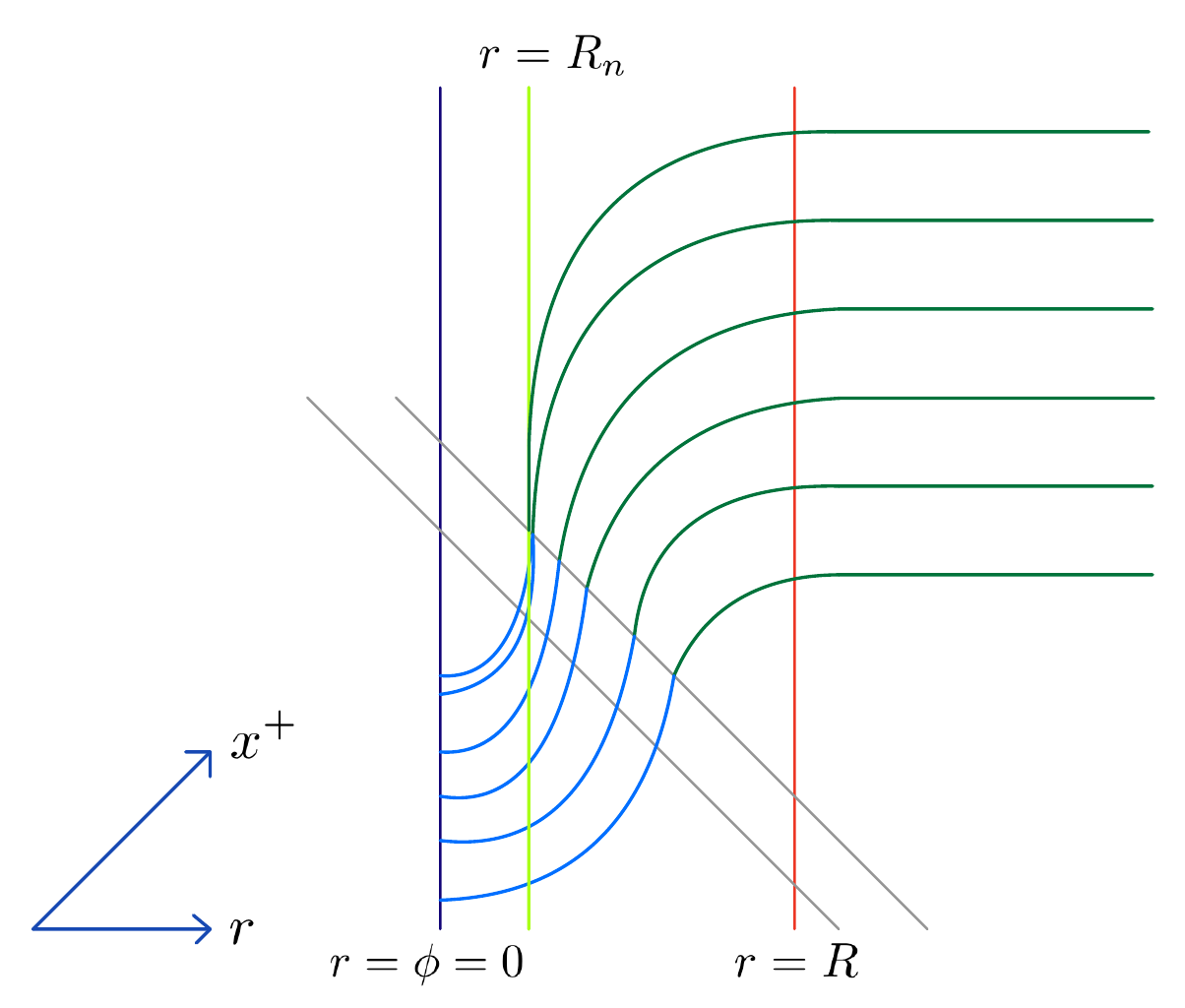}
\end{center}
\caption{A slicing of the 2d BH spacetime which in the post-formation region utilizes nice slices.  These asymptote to the radius $R_n<R$, shown in light green.  However, in the pre-formation region these would not give Cauchy slices; Cauchy slices can be constructed by extending the slices differently into the early region, as shown in light blue.}
\label{FigNS}
\end{figure}

The formula \eqref{slicedef} unifies various different descriptions of the spacetime\cite{SEHS}.  Using $x^+$ as time corresponds to $S=0$.  Schwarzschild time slices  correspond to $S(r)=r_*(r)$.  Another simple choice is that of ``straight slices," with $S(r)=r$. A useful case intermediate between these is to take $S(r)$ to be a monotonic function that asymptotes to minus infinity for some finite value $r=R_n<R$; for sufficiently large $R_n>0$, these are ``nice" in that they avoid strong coupling\cite{Waldnice,LPSTU}. The straight slice parameterization and others with $S\rightarrow -\infty$ as $r\rightarrow-\infty$ can be extended into the preformation vacuum region to provide Cauchy slices for the full spacetime, as shown for example in Fig.~\ref{FigEF}.  While the nice slices provide Cauchy slices for the post-formation vacuum BH, they in contrast do not provide Cauchy slices for the earlier vacuum region.  But, one can match these nice slices to another slicing at earlier times to find corresponding Cauchy slices, as shown in Fig.~\ref{FigNS}.  

\begin{figure}[!hbtp] \begin{center}
\includegraphics[width=11cm]{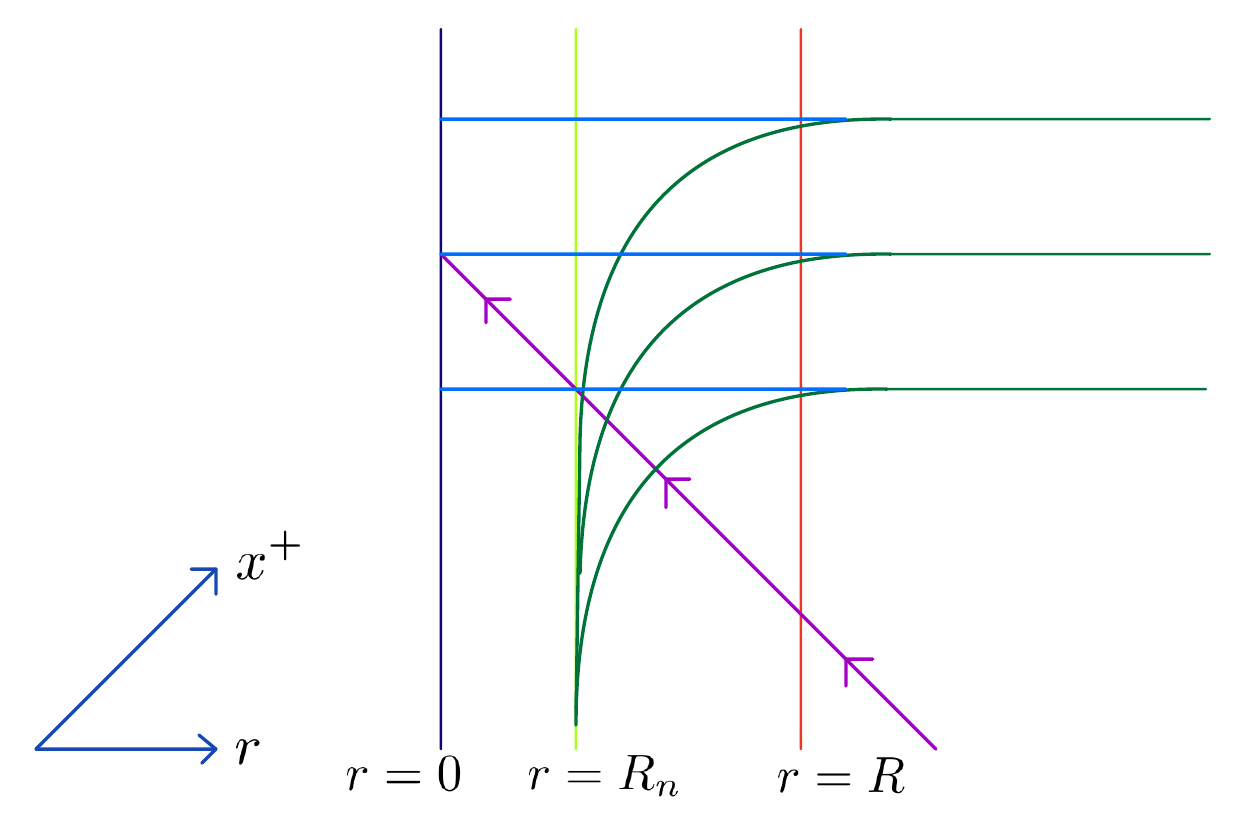}
\end{center}
\caption{Slices in a four (or higher) dimensional Schwarzschild BH spacetime. Straight slices end at $r=0$, and so do not provide good Cauchy slices; for example, data corresponding to the infalling excitation shown does not register on later straight slices, unless additional degrees of freedom are included ``at $r=0$".  Slices that have nice behavior, in that they avoid the singularity, as shown, do in contrast provide Cauchy slices.}
\label{Fig4d}
\end{figure}

Notice that higher-dimensional spacetimes behave differently; for example higher-dimensional Schwarzschild terminates at $r=0$.  This means that in higher dimensions straight slices and other similar choices which intersect $r=0$ do not give Cauchy slices for the spacetime, as shown in Fig.~\ref{Fig4d}.  So, in higher dimensions one needs nice slice like behavior if one wants to describe evolution on Cauchy slices, or alternately one needs some other completion of the Hilbert space to describe the ``missing states" at $r=0$.

With a general time translation invariant slicing \eqref{slicedef}, the  BH metric \eqref{Vaidform} takes the ADM form \eqref{ADMmet}, with\cite{NVU}
\beq\label{ADMvar}
q=q_{rr}={S'[2-f(x^+,r)S']} = \frac{1}{N^2}\quad,\quad N_r=1-f(x^+,r)S'\ 
\eeq
and $f(x^+,r)$ given in \eqref{fdef}.  One may alternately use a more general spatial parameter $x(r)$ as a coordinate along the slices, and transform these expressions accordingly.

\subsection{Schr\"odinger evolution of the state}

\subsubsection{Schr\"odinger description of the state}

One goal of this paper is to provide an explicit description of some features of the evolving quantum state, in a Schr\"odinger picture.  The evolution can be described with respect to a particular basis for the Hilbert space, which arises from making some choices.\footnote{In the end physical quantities should be independent of such choices.  We will leave more complete discussion of this statement, which is closely related to the gauge invariance of gravity, for future work.}  First is a choice of time slicing, as for example in the preceding discussion.  Second is a choice of coordinates along the slices.  Third, one needs to choose a basis of modes, with different choices useful in different contexts.   In a Schr\"odinger description, a mode basis can be defined by specifying  
 a complete set of pairs of functions $\gamma_i(x)= (f_i(x), \pi_i(x))$ on  a general time slice, with a complex structure\cite{More,Kay,AgAs} that separates these into the ``positive frequency" modes $\gamma_A(x)= (f_A(x), \pi_A(x))$ and conjugate ``negative frequency" modes $\gamma_A^*(x)$.  On a $t=$constant slice, one then expands the field operators as
\beq\label{opexp}
F(x)=\sum_A\left[ a_A f_A(x) + a_A^\dagger f_A^*(x)\right]\quad ,\quad \Pi(x)=\sum_A\left[ a_A\pi_A(x) + a_A^\dagger \pi_A^*(x)\right]\ .
\eeq
If the mode basis is normalized such that
\beq\label{modenorm}
{(\gamma_A,\gamma_B) =  \delta_{AB}\quad,\quad (\gamma_A,\gamma_B^*)=0\ ,}
\eeq
with the norm
\beq\label{normdef}
{(\gamma_1,\gamma_2)=i\int dx \sqrt q (f_1^* \pi_2 - \pi_1^*f_2)}
\eeq
then, from the canonical commutators \eqref{CCR}, the operators $a_A,a_B^\dagger$ satisfy 
\beq\label{acomms}
{[a_A,a_B^\dagger]=\delta_{AB}\quad,\quad[a_A,a_B]=[a_A^\dagger,a_B^\dagger]=0\ .}
\eeq
Basis states for the Hilbert space follow from the standard Fock construction, with creation operators $a_A^\dagger$ acting on a vacuum $|0\rangle$ annihilated by the $a_A$.  

Different choices of time slicings, spatial coordinates, and modes yield  different bases for the Hilbert space.
 For example, as discussed in \ref{Params}, once a time slicing is chosen, the coordinates $-x^-$, $-X^-$, $r$, or a more general regular $x(t,r)$, are all regular across the horizon.  With a general such regular coordinate $x$, one useful choice of modes follows by taking
\beq\label{pwmodes}
f_k= e^{ik x}\quad  ,\quad \tilde f_k = e^{-ik x}\ ,
\eeq
for arbitrary $k$,
and then defining $\pi_k$, $\tilde \pi_k$ by requiring that $f_k$ and $\tilde f_k$ extend to right and left moving solutions, annihilated by $\partial_\pm$, respectively.  Due to the regularity of $x$ at the horizon, such bases are naturally  referred to as {\it regular bases}\cite{SEHS}.
The field operator can then be taken to have expansion
\beq\label{Fex}
F(x) = \int_0^\infty \frac{dk}{\sqrt{2\pi} 2 k} (a_k e^{ikx} + {\tilde a}_k e^{-ikx} + {\rm h.c.})\ ,
\eeq
and similarly for $\Pi(x)$, 
where $a_k$, $\tilde a_k$ are right/left moving operators with  commutator normalization
\beq
[a_k,a^\dagger_{k'}]=4\pi k\delta(k-k')=[\tilde a_k,\tilde a^\dagger_{k'}]\ .
\eeq

In such a Schr\"odinger picture description, the operators $F(x)$ and $\Pi(x)$ of \eqref{opexp} are time independent, and evolution is that of the state, as in \eqref{stateevol}.  The initial state for the BH (excluding the infalling matter) is most simply described in terms of the modes $\exp\{\pm ikr\}$ in the asymptotic past.  If we use straight slices, with $x^+=t+r$, in this region,  the initial state is just the vacuum $|0\rangle_-$ for these modes.  The description of the subsequent post-formation evolution of the state then depends on the choices we have made of slicing, coordinates, and basis.

\subsubsection{Hamiltonians, pictures, and evolution}

We have found that, for a general choice of slices, with coordinates $(t,x)$, the  leading-order evolution between slices is given by the ADM matter hamiltonian, 
from \eqref{HF},
\beq\label{Fham}
H=\frac{1}{4\pi}\int dx \sqrt q \left\{N \left[(2\pi\Pi)^2 + \frac{ (\partial_x F)^2}{q}\right] + 4\pi N^x \Pi \partial_x F \right\}\  ,
\eeq
with canonical commutators \eqref{CCR}.
Notice that this is readily extended to include interaction terms.
This hamiltonian depends on the lapse $N$ and shift $N^x$, which are specified by giving the vector
\beq
\xi=(\partial/\partial t)_x\quad \leftrightarrow\quad \xi^\mu=(\partial x^\mu/\partial t)_x\ 
\eeq 
connecting corresponding points on nearby slices (here the subscript denotes the variable held fixed under differentiation).  
It is also useful to rewrite $H$, using the expressions \eqref{Fmom} and \eqref{ndef} for $\Pi$ and the unit normal $n^\mu$, as
\beq\label{HT}
H=H_\xi= \frac{1}{4\pi} \int dx \sqrt q \left[\frac{(\partial_t F)^2}{2N} + \frac{N}{2} g^{xx}\partial_xF\partial_x F\right] = \frac{1}{\pi}\int dx\sqrt q\ n^\mu \xi^\nu T_{\mu \nu}\ .
\eeq

Here we introduce the notation $H_\xi$ to emphasize the dependence of
 the hamiltonian not only on the slicing but also on the choice of coordinate along the slices:
the vector $\xi$ is defined with fixed $x$, and so depends on the choice of $x$.\footnote{This dependence is also directly seen in alternate approaches, such as the covariant canonical quantization; see {\it e.g.} \cite{DoGi2} for a brief overview and further references, or \cite{HaWu}.}   A key point is that different choices of slices and coordinate lead to different hamiltonians.
One might say that these choices define {\it different 
Schr\"odinger pictures}.  

In the special case where $\xi^\mu$ is a Killing vector, like in the post-formation region with coordinate $x$ taken to be a function only of $r$, conservation of $T_{\mu\nu}$ then implies that $H_\xi$ is conserved.

In terms of chiral components, using $T_{+-}=0$,
the hamiltonian \eqref{HT} can be expressed as 
\beq\label{HTLR}
H_\xi=\frac{1}{\pi}\int dx \sqrt q \left( n^- \xi^- T_{--} + n^+ \xi^+ T_{++}\right)\ ,
\eeq
For example if we consider a stationary slicing given by \eqref{slicedef}, for some $S(r)$, that definition, and the coordinate relation \eqref{basicc}, give the relations of the chiral coordinates $x^\pm$ to $t,r$.  

Focussing first on left (ingoing) movers, one easily finds from the normal \eqref{ndef} and the ADM expressions \eqref{ADMvar} that $n^+=S'/\sqrt{q_{rr}}$.  For a spatial coordinate $x(r)$ with no extra time dependence, 
$\xi^+=1$, and the left-moving hamiltonian becomes
\beq
H_\xi^L=\frac{1}{2\pi} \int \frac{dx}{ \left(\partial x^+/ \partial x\right)_t } \partial_x F\partial_x F\ 
\eeq
Specifically, with  straight slices $S(r)=r$ and spatial coordinate $x=r$, the evolution is the same as in a flat background,
\beq\label{LHam}
H_\xi^L= \int_0^\infty \frac{dk}{4\pi k} k\,  \tilde a^\dagger_k \tilde a_k\quad  +\quad {\rm constant}\ .
\eeq

The right (outgoing) movers behave more nontrivially.  With a general choice of spatial coordinate $x(r,t)$ or $x(x^+,x^-)$, one finds from a short calculation using \eqref{basicc} the needed expressions
\beq
\sqrt{q_{xx}} n^-=-\left(\frac{\partial x^-}{\partial x}\right)_t
\eeq
and $\xi^-= (\partial x^-/\partial t)_x$.  Substituting the right-moving part of the evolving solution from \eqref{Fex} into the hamiltonian \eqref{HTLR} then gives
\beq\label{Hm}
H_\xi^R=\frac{1}{2\pi}\int dx\, \gamma(t,x) \partial_x F\partial_x F =\int_0^\infty \frac{dk}{4\pi} \frac{dk'}{4\pi}\left[ A(k,k') a_k^\dagger a_{k'} + B(k,k') a_k^\dagger a_{k'}^\dagger +{\rm h.c.}\right]\ ,
\eeq
with the function
\beq
\gamma(t,x)=-\left(\frac{\partial x^-}{\partial t}\right)_x\Big/ \left(\frac{\partial x^-}{\partial x}\right)_t 
\eeq
determined by the slice parameterization, and with time-dependent coefficients 
\beq\label{ABdefs}
A(k,k') = \int dx \gamma(t,x) e^{-i(k-k') x}\quad ,\quad 
B(k,k') = -\int dx \gamma(t,x) e^{-i(k+k') x}\ .
\eeq
The hamiltonian \eqref{Hm}  governs the evolution in a regular basis \eqref{Fex}.  The presence of the coefficients $B(k,k')$  describes particle creation in this basis.  If one instead begins with an interacting theory, one also expects terms that are cubic or higher-order  in the ladder operators.

For examples of this general evolution, consider a slicing defined by the stationary definition \eqref{slicedef}, extended into the pre-formation region.  
For coordinate along (within) the slices, one choice is to use
$x^-$, and it is then natural to use  the ``conformal" modes $f_k=e^{-ik x^-}$ for the right-moving basis.
Then in the $(t,x^-)$ coordinates $\xi^\mu=(1,0)$, so $\gamma\equiv0$, the hamiltonian $H_\xi^R=0$, and in this picture evolution is trivial.  The same follows if the spatial coordinate is the Kruskal $X^-$ of \eqref{Krusk}.

In contrast, for evolution in a picture with spatial coordinate $x=r$, we instead find, from the coordinate relations \eqref{basicc}, \eqref{slicedef},
\beq\label{gammatr}
\gamma(t,r) = \frac{f(x^+,r)}{2-f(x^+,r)S'}\ ,
\eeq
with $f(x^+,r)$ given by \eqref{fdef}, generalizing a result in \cite{SEHS}. For a given slicing determined by $S(r)$, the initial state $|0\rangle_-$ evolves to a nontrivial excited state, with a basis-dependent description in each of these cases determined by these expressions and evolution by the hamiltonian \eqref{Hm}.  

The differences between the nontrivial $H_\xi$ in this case and the preceding $H_\xi=0$  illustrates the non-trivial role of the choice of spatial coordinates in determining $H_\xi$, and the different pictures.

Some properties of the state are more clearly exhibited by expanding the field in a different basis,
defined in terms of the coordinates $y^-$, $\hat y^-$ of \eqref{ydef}, \eqref{yhatdef}:
\beq\label{Emodes}
f_\omega=e^{-i\omega y^-}\quad ,\quad \hat f_\omega=e^{-i\omega \hat y^-}\ .
\eeq
In the static post-formation region, these are eigenstates of the translation operator $\partial_+$.  
If one considers evolution defined using coordinates $y^-$, $\hat y^-$ along the slices, the hamiltonian vanishes, as above.  However, the mode basis \eqref{Emodes} may also be used to describe the evolution with respect to spatial coordinate $x=r$, so that $\xi=\partial/\partial t\vert_r$.  The modes \eqref{Emodes} are rewritten in the $r,t$ coordinates using the definitions \eqref{ydef}, \eqref{yhatdef} of $y^-$, $\hat y^-$, the coordinate relation \eqref{basicc}, and the slice definition \eqref{slicedef}.  Alternately, the corresponding hamiltonian may be reexpressed as
\beq\label{Hym}
H_\xi^R= \frac{1}{2\pi} \int_{-\infty}^\infty dy^- \left(-\gamma \frac{\partial y^-}{\partial r} \right)\left(\partial_-F\right)^2 + \frac{1}{2\pi} \int_{-\infty}^{Y^-} 
d \hat y^-  \left(-\gamma \frac{\partial \hat y^-}{\partial r} \right)\left(\hat \partial_-F\right)^2\ .
\eeq
where $Y^-$ is the value of $\hat y^-$ corresponding to $r=-\infty$, and where one easily shows
\beq
-\gamma \frac{\partial y^-}{\partial r}=\frac{\partial y^-}{\partial t}=1+[\Delta-\Delta(x^+)]e^{y^-}\quad ,\quad -\gamma \frac{\partial \hat y^-}{\partial r}=\frac{\partial \hat y^-}{\partial t} =-1+[\Delta-\Delta(x^+)]e^{-\hat y^-}\ .
\eeq
From these last expressions one sees that in the post formation region $x^+>x_f^+$, where $\Delta(x^+)=\Delta,$ the hamiltonian is of the standard form for flat space modes, with an extra minus sign for the internal modes, generalizing the corresponding result in \cite{SEHS}.

While in the post-formation regime, the modes \eqref{Emodes} conveniently diagonalize the hamiltonian, they are of course singular; moreover, their singular behavior is ``teleological" in that it depends on the final location of the event horizon.  Aspects of this singular behavior are sometimes taken literally,  but we see that it is more plausible to regard such aspects as basis-dependent artifacts, particularly in view of the corresponding description in regular bases.

\subsubsection{Evolution of the Hawking state: transitory behavior}

We now turn to describing  behavior of the evolving state.  We assume that in the pre-formation region the state is the Minkowski vacuum.  Once the BH forms, the state develops non-trivial time dependence and we refer to it as the {\it Hawking state}.

In the region before the incoming shock, this state is obviously most easily described using the slicing \eqref{slicedef} with $S(r)=r$.  For spatial coordinate, one may use either $r$ or $x^-$.  The incoming vacuum state is $|0\rangle_-$, which behaves identically to $|0\rangle_r$ inside the shock; these states are the vacua annihilated by the corresponding positive frequency operators.  

As above, the different choices of spatial coordinate lead to different formulas for the hamiltonian in this region, giving a very simple illustration of the different pictures.  With spatial coordinate $r$, evolution is via a hamiltonian of standard flat-space form.  With picture based on spatial coordinate $x^-$, we have $\xi^-=0$, and by \eqref{HTLR} the hamiltonian for outgoing modes vanishes.  

If, alternately, nontrivial slices  given by a more general $S(r)$ are chosen, the evolution of the state looks more complicated, as follows from the general form of the hamiltonian \eqref{Hm}, \eqref{ABdefs}, and \eqref{gammatr}.
For a general choice of slices and coordinates, the  time dependence of the Hawking state is given by
\beq
|\Psi_H(t,\xi)\rangle = e^{-i\int_{-\infty}^t dt' H_\xi(t')} |0\rangle_-\ ,
\eeq
where the hamiltonian for right movers is as in \eqref{Hm}.  
\begin{figure}[!hbtp] \begin{center}
\includegraphics[width=14cm]{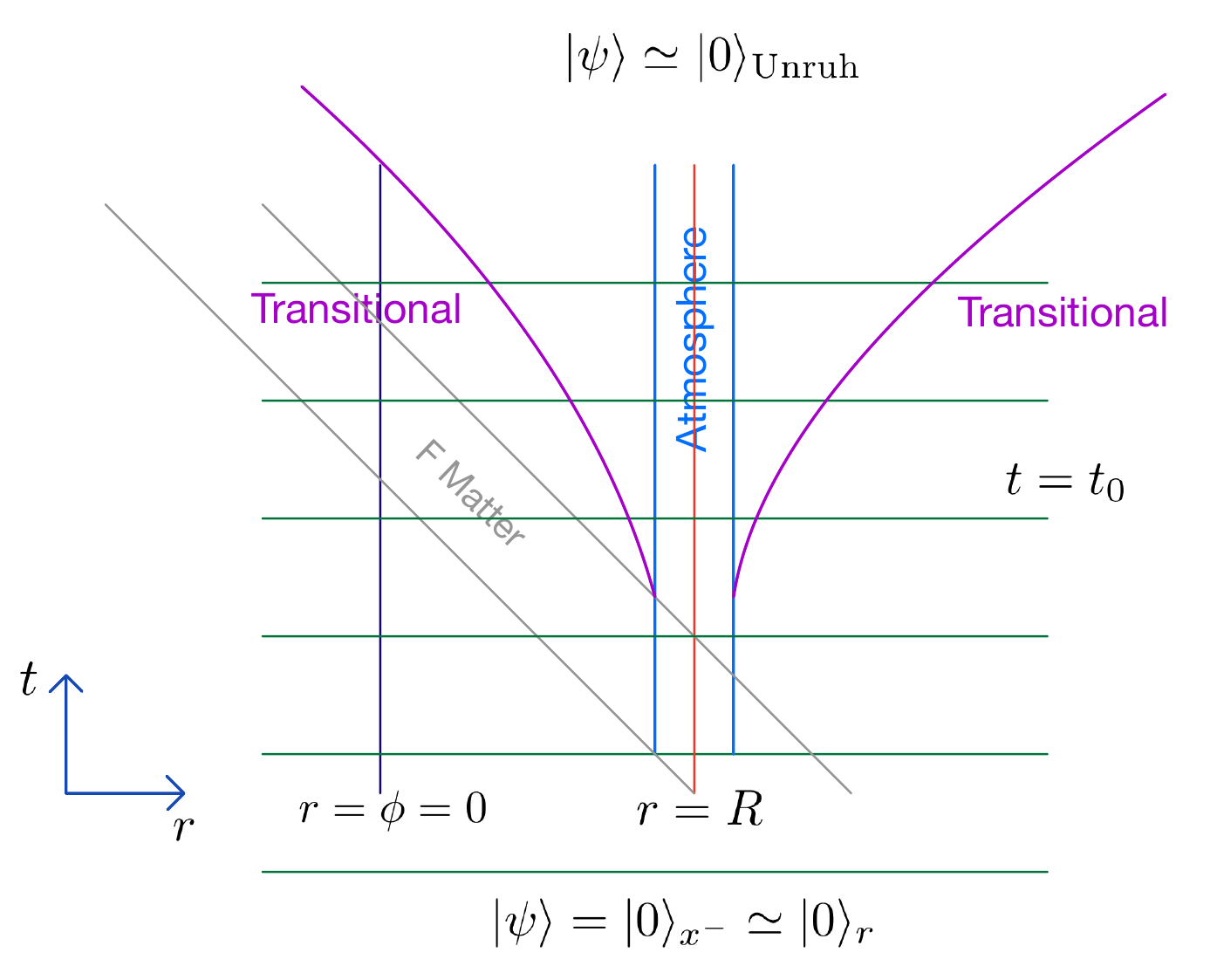}
\end{center}
\caption{A depiction of the evolving state in the background geometry induced by an incoming $F$-wave, which produces a BH.  A straight time slicing is shown.  In the region after the incoming wave turns off, the classical solution is static.  The induced quantum state of the matter, with initial vacuum state, has a transitional region, resulting from the time-dependence of the collapse. This time-dependent radiation escapes to infinity or falls into the strong-coupling region, after which the state
asymptotically settles down to a stationary state.  This late-time state is approximated by the Unruh vacuum.  The outgoing Hawking excitations, and their inside partners, can be thought of as produced in an atmosphere region $|r-R|\lesssim 1$.  }
\label{Figregions}
\end{figure}

As the infalling matter creates the BH, outside observers will see Hawking radiation produced. 
As we have described,  slicings of the stationary form \eqref{slicedef} simplify the description in the post-formation region.  Here, too, one may use modes based on spatial coordinate $x^-$ or $r$, or make another choice, and determine the evolution from the hamiltonian \eqref{Hm}.  A particularly simple choice is to maintain the straight slicing, $S(r)=r$, into this region, and use spatial coordinate $r$.  Such a slicing is pictured in Fig.~\ref{Figregions}.  With $f(x^+,r)$ given in \eqref{fdef}, and with the definition 
\beq
R(x^+)=\hf\log M(x^+)\ ,
\eeq
we find
\beq\label{gstraight}
\gamma(t,r)= \frac{f}{2-f} = \tanh[r-R(x^+)]\ .
\eeq
This transitions smoothly from the initial value $\gamma=1$ (flat hamiltonian) to that of the final BH solution \eqref{Vaidform} with constant $R$.  

Much of the behavior of the evolving state can also be inferred from the relationship between the different coordinates.  The coordinate relations become, combining 
\eqref{basicc}, \eqref{ydef}, \eqref{yhatdef}, \eqref{Krusk}, and using the straight slicing\footnote{For more general slices, the left hand side of this expression is replaced by $2\sinh[r-R(x^+)] e^{-[t-R(x^+)]+r-S(r)} +\Delta(x^+) $.} $x^+=t+r$,
\beq\label{LTcoord}
2\sinh[r-R(x^+)] e^{-[t-R(x^+)]} +\Delta(x^+) = e^{-x^-}  = -X^-+\Delta = \begin{cases} e^{-y^-}+\Delta \ ,\ r>R\\ -e^{\hat y^-}+\Delta \ ,\ r<R\end{cases}\ .
\eeq

In the picture based on spatial coordinate $x^-$, the state remains simply $|0\rangle_-$.  In a picture based on spatial coordinate $r$, the infalling matter produces nontrivial excitations.  
If one considers a time slice just after the infalling matter has crossed the horizon, as in Fig.~\ref{Figregions}, there will be transitional excitations that escape to $r=\infty$ or travel towards the singularity at  $r=-\phi=-\infty$.  As the state evolves forward in time, Hawking quanta will then continue to be produced by the dynamics of the hamiltonian \eqref{Hm}.  As we will describe further, the long-time evolution is governed by the near-horizon behavior of the state, which is vacuum-like at short distances.  Near the horizon, at $|r-R|\ll 1$, this vacuum-like behavior may be described by specifying the vacuum for the $r$ modes, the $x^-$ modes, or the $X^-$ modes, since they are linearly related in this limit.  The difference between these states, and their difference with the Hawking state, is in excited radiating modes in the regions $|r-R|\gtrsim 1$.  This description, and the additional discussion of this paper, lends further support to
earlier arguments\cite{SGBoltz} that the quanta of Hawking radiation emerge from a ``quantum atmosphere," rather than a much thinner region near the horizon;  in the 2d context this is the region  with $|r-R|\lesssim 1$ and has a thickness comparable to the wavelength of the Hawking modes.

\subsubsection{Evolution of the Hawking state: long-time behavior}

For simplicity, we can choose the end of the collapse to be $x_f^+=0$.  To describe the long-time behavior of the radiation, consider a slice with $t\gg R$.  On this slice, the transitory radiation has escaped to far from the BH or fallen deep into the strong-coupling region, as shown in Fig.~\ref{Figregions}, and will be ignored.  In this limit, the BH metric, {\it e.g.} \eqref{Vaidform}, has the time translation symmetry $x^+\rightarrow x^+ + const;$  the coordinate relations \eqref{LTcoord} correspondingly simplify, with $R(x^+)=R$ and $\Delta(x^+)=\Delta$.
In particular notice that the $\sinh$ leads to a nice interpolation where the $r$ coordinate is proportional to Kruskal $X^-$ near the horizon, and 
linear in the good asymptotic coordinates $-y^-$ or $\hat y^-$ far from the horizon.  

Initial data to determine the state may be specified on a slice at a $t_0$ just after horizon crossing.  For the purposes of discussing the 
 long-time radiation in a finite region near the BH, different regular initial states may be chosen on this slice.  These include the Hawking state, 
 the Unruh vacuum, which is the vacuum $|0\rangle_{X^-}$ associated with the modes $e^{-i\omega X^-}$, or the vacuum $|0\rangle_r$ associated with the modes $e^{ikr}$.  While these differ in their excitations for $|r-R|\gtrsim 1$ at time $t_0$, these differences  dissipate leaving the same long-time local behavior of the states in an expanding region near the BH, shown as the region above the transitional regions in   
Fig.~\ref{Figregions}.

The evolution of such a regular state in a picture and basis \eqref{pwmodes} given by spatial coordinate $x=r$  is governed by the hamiltonian \eqref{Hm}, with $\gamma$ given by
\eqref{gstraight} with constant $R$; this limit gives the coefficients\cite{SEHS}
\beq
B(k,k') = \frac{i\pi}{\sinh[\pi(k+k')/2]}e^{-i(k+k')R}\ .
\eeq
 As pointed out in \cite{SEHS}, these both exhibits the expected thermal spectrum, $B(k,k')\propto \exp\{-\pi(k+k')/2\}$, with temperature $T=1/(2\pi)$, and show that excitations are produced at wavelengths with $k=\calo(1)$, rather than at much shorter wavelengths.

While the  evolution with hamiltonian \eqref{Hm}  is regular and exhibits these key features, it is rather complicated.  This suggests a change of basis to analyze other features of the evolving state.  Specifically, we found that the hamiltonian can be diagonalized by using the mode basis $e^{-i\omega y^-}$ and $e^{-i \omega \hat y^-}$, which in $r,t$ coordinates become, from \eqref{LTcoord},
\bea
e^{-i \omega y^-} &= [2\sinh(r-R)]^{i\omega} e^{-i\omega(t-R)}\quad,\quad r>R\cr
e^{-i \omega \hat y^-} &= [2\sinh(R-r)]^{-i\omega}e^{i\omega(t-R)}\quad,\quad r<R\ ,
\eea
and will be taken to have associated annihilation operators $b_\omega$, $\hat b_\omega$ (see {\it e.g.} the expansion \eqref{Freln}).  In the long-time limit, which 
sets $Y^-=\infty$ in \eqref{Hym}, that hamiltonian then takes the simple form
\beq\label{Hb}
H_\xi^R=\int_0^\infty \frac{d\omega}{4\pi\omega} \omega (b_\omega^\dagger b_\omega - \hat b_\omega^\dagger \hat b_\omega)\ 
\eeq
(up to a normal-ordering constant, which formally cancels). Of course, the  simplicity of this expression is belied somewhat by the singular behavior of the mode basis, and its teleological behavior near the horizon.

To exhibit the evolving state in this $b, \hat b$ basis, we need to rewrite the initial state at $t_0$ in this basis.  For simplicity, begin by considering the evolution of the state $|0\rangle_r$ from some initial time $t_0\gg R$, to ignore the transitional region.  
The necessary Bogolubov transformation between bases is found by equating the expansion \eqref{Fex} of $F$ in $r$ modes to the corresponding expansion in $y^-,\hat y^-$ modes (see Appendix \ref{BOG}), and the Bogolubov coefficients are simply the Fourier coefficients of the latter mode functions, 
\bea
 \theta(r-R)e^{-i\omega y^-} &=& \int_0^\infty dk (\alpha_{\omega k}^+ e^{ikr} + \alpha^-_{\omega k} e^{-ikr})\cr
 \theta(R-r)e^{-i\omega \hat y^-}  &=& \int_0^\infty dk (\hat \alpha_{\omega k}^+ e^{ikr} + \hat \alpha_{\omega k}^- e^{-ikr})\ .
\eea
These give
\beq\label{aint}
\alpha_{\omega k}^\pm = \frac{1}{2\pi}  \int_R^\infty dr e^{-i\omega y^- \mp ikr}= \frac{e^{-i\omega(t-R)\mp i k R}}{2\pi} \int_0^\infty dx e^{i\omega \ln(2\sinh x) \mp ikx}
\eeq
and
\beq\label{ahint}
\hat \alpha_{\omega k}^\pm = \frac{1}{2\pi}  \int_{-\infty}^R dr e^{-i\omega \hat y^- \mp ikr} 
= \frac{e^{i\omega(t-R)\mp i k R}}{2\pi} \int_0^\infty dx e^{-i\omega \ln(2\sinh x) \pm ikx}\ ,
\eeq
where the coordinate relations  \eqref{LTcoord} and simple changes of variables $x=\pm(r-R)$ have been used.
These can be explicitly computed (see Appendix \ref{BOG}) in terms of beta functions, but some of their features are most apparent from these integral expressions.

The state $|0\rangle_r$ may then be formally written\cite{GiNe}
\beq \label{Bogstate}
|0\rangle_r\ ``\propto"\ \exp\left\{ -\hf (b^\dagger\ \hat b^\dagger) {\cal B}^*{\cal A}^{-1}\begin{pmatrix} b^\dagger\\ \hat b^\dagger\end{pmatrix}\right\} |\hat 0, 0\rangle \ .
\eeq
where $|\hat 0, 0\rangle$ denotes the product vacuum $|\hat 0\rangle\otimes |0\rangle$ and where $\cal A$ and $\cal B$ are the (infinite dimensional) matrices\footnote{The matrix multiplication convention is exemplified by the expression $\left[(b^\dagger\ \hat b^\dagger) {\cal B}^*\right]_k = \int \frac{d\omega}{4\pi\omega}\left( b^\dagger_\omega \alpha^{-*}_{\omega k} + \hat b^\dagger_\omega \hat \alpha^{-*}_{\omega k} \right)$.}
\beq
{\cal A}=\begin{pmatrix}\alpha^+_{\omega k}\\ \hat \alpha^+_{\omega k}\end{pmatrix}\quad ,\quad {\cal B}=\begin{pmatrix}\alpha^-_{\omega k}\\ \hat \alpha^-_{\omega k}\end{pmatrix}\ .
\eeq
This expression is only formal, due to the infinite product over modes at arbitrarily large $\omega$; the state $|0\rangle_{r}$ does not actually lie in the Fock space built by exciting $|\hat 0\rangle\otimes | 0\rangle$.\footnote{This, in turn, is associated with the Type-III property of operator algebras in QFT; see, {\it e.g.}, \cite{Haag}.}  In particular, the hamiltonian $H_\xi$ of \eqref{HT}, \eqref{Hb} is infinite on this expression for $|0\rangle_r$.  A finite hamiltonian is defined by subtracting a normal ordering constant.  

Since there is different behavior in the regions $|r-R|\ll 1$ and $|r-R|\gg 1$, studying this behavior requires localization, which is not directly achieved in plane wave bases.  One approach is to instead use a wavepacket basis, by  introducing a basis of functions $f_i(x) =\int dk f_i(k) e^{ikx}$ that for example have compact support in $x$ (or $r$).  A somewhat simpler basis  is the wavepacket basis used in 
\cite{Hawk}\cite{GiNe}, with mode functions
\beq
f_{jn}(x) = \epsilon^{-1/2} \int_{j\epsilon}^{(j+1)\epsilon} \frac{dk}{\sqrt{2k}} e^{ik(x-2\pi n/\epsilon)}\ ,
\eeq
for some small $\epsilon$, and with integer $j,n$ and $j\geq0$.  
Though these modes do not have compact support in position, they do have rapid falloff with distance, and capture important aspects of the localization.
We will focus on some general aspects that are expected to result from these or other localization procedures.  In particular, the expression \eqref{Bogstate} may be extended to such a basis.

From \eqref{Bogstate}, we see that the Bogolubov coefficients $\alpha_{\omega k}^-$, $\hat \alpha_{\omega k}^-$ describe excitations in the $b,\hat b$ basis.  In the regions with $|r-R|\gg 1$, the coordinate relation \eqref{LTcoord} becomes
\beq\label{Asympy}
y^-= t-r+\calo\left(e^{-2(r-R)}\right)\ , \ r-R\gg1\quad ;\quad \hat y^- = 2R-r-t+\calo\left(e^{2(r-R)}\right)\ ,\ R-r\gg 1\ ,
\eeq
showing that the $r$ modes and the $y^-,\hat y^-$ modes are identical, up to exponentially small terms in $R-r$, and so these coefficients are effectively vanishing in these regions.   This means that for excitations in regions far from the horizon, the initial $t=t_0$ state $|0\rangle_r$ behaves as the vacuum $|\hat 0,0\rangle$.

On the other hand, the state $|0\rangle_r$ near the horizon is non-trivially excited in the $b, \hat b$ basis, as follows from the near horizon limit of \eqref{aint}, \eqref{ahint}.  
A  simple illustration of this arises from using the fact that in 
the near-horizon limit, $r$ and $X^-$ are linearly related, so the near-horizon structure of the vacuum $|0\rangle_r$ and the  Unruh vacuum $|0\rangle_{X^-}$ are the same.\footnote{This may be made more precise by arguments using the wavepackets we have described\cite{GiNe}.}  Evolution of the initial state $|0\rangle_{X^-}$, from $t=t_0$ is particularly simple.  While at $t=t_0$ it differs nontrivially from $|0\rangle_r$ for $|r-R|\gtrsim 1$, the boundary of the region where they differ goes to $r=\pm\infty$, analogously to Fig.~\ref{Figregions}, so at later times they have the same behavior in a large region containing the horizon.

The simplification for $|0\rangle_{X^-}$  arises from an analyticity argument \cite{Unru,Hawk-incoh}\cite{GiNe}  relating the vacuum $|0\rangle_{X^-}$ to that of the energy eigenmodes.
Through analytic continuation we find from the coordinate relations \eqref{LTcoord}
\beq
y^- = -\ln(-X^-) = - \hat y^- \pm i\pi \ .
\eeq
If we place the branch cut in the upper half $X^-$ complex plane, to maintain analyticity in the lower half plane, that describes a positive frequency prescription for the function $(-X^-)^{i\omega}$, which then limits to the positive frequency expression
\beq
u_\omega^1\propto (-X^-)^{i\omega}\rightarrow e^{-i\omega y^-} \theta(-X^-) + e^{-\pi\omega} e^{i\omega \hat y^-} \theta(X^-)\ 
\eeq
on the real axis.  The same analytic/positive frequency prescription for $(X^-)^{-i\omega}$ results in 
\beq
u_\omega^2\propto(X^-)^{-i\omega} \rightarrow e^{-i\omega \hat y^-} \theta(X^-) + e^{-\pi\omega} e^{i\omega y^-} \theta(-X^-)\ .
\eeq
The field $F$ may be expanded in the modes $u_\omega^1$, $u_\omega^2$ and their conjugates, analogously to \eqref{opexp}.  
Since the modes $u_\omega^1$, $u_\omega^2$ are positive frequency, the corresponding operators 
annihilate 
$|0\rangle_{X^-}$, 
\beq\label{stateann}
a^1_\omega |0\rangle_{X^-}= \frac{b_\omega - e^{-\pi\omega}\hat b^\dagger_\omega}{\sqrt{1-e^{-2\pi\omega}}}|0\rangle_{X^-}=0 \quad,\quad a^2_\omega|0\rangle_{X^-}=\frac{\hat b_\omega - e^{-\pi\omega} b^\dagger_\omega}{\sqrt{1-e^{-2\pi\omega}}}|0\rangle_{X^-}=0\ .
\eeq
  This implies that 
\beq
0=(a^{1\dagger}_\omega a^1_\omega -a^{2\dagger}_\omega a^2_\omega)|0\rangle_{X^-} = (b^\dagger_\omega b_\omega - \hat b^\dagger_\omega \hat b_\omega)|0\rangle_{X^-}\ ,
\eeq
and so there is a precise pairing between occupation of internal and external excitations.  This, together with \eqref{stateann}, can then be solved (see, {\it e.g.}, \cite{GiNe}) to find the formal expression
\beq\label{pairstate}
|0\rangle_{X^-}\ ``="\ C \sum_{\{n_\omega\}} e^{-\pi \int d\omega \omega n_\omega }| {\{\widehat {n_\omega}\}}\rangle|\{n_\omega\}\rangle\ .
\eeq
in terms of occupation number states $|\{n_\omega\}\rangle$, $|\{\widehat {n_\omega}\}\rangle$ for the $b_\omega$, $\hat b_\omega$ operators, with $C$ a constant.  

This provides a simple example of the more general expression \eqref{Bogstate}.
Once again it  is only formal, due to the infinite product over modes at arbitrarily large $\omega$; the states $|0\rangle_r$  or $|0\rangle_{X^-}$ do not actually lie in the Fock space build by exciting $|\hat 0\rangle| 0\rangle$.  In particular, the hamiltonian $H_\xi$ of \eqref{HT} is infinite on this expression for $|0\rangle_{X^-}$ and is only finite after subtracting a normal ordering constant.  The expressions may also be regulated by imposing a cutoff on the the frequency $\omega$.  

The expression \eqref{pairstate} may be extended to an analogous one for localized wavepackets\cite{GiNe}, and then captures the behavior of $|0\rangle_r$ or $|0\rangle_{X^-}$ for the near-horizon modes in the region $|r-R|\ll 1$.  As noted above, with $|0\rangle_r$ the modes $b,\hat b$ in the region $|r-R|\gg 1$ are approximately in their vacuum. Also as noted, other regular states will have the same pairing property of their ultraviolet excitations, but will differ in their excitation in this far region.

The  different descriptions of the initial state at $t=t_0$ then evolve in time.  In view of the simple form of the hamiltonian \eqref{Hb}, the evolution is simplest to give in the $b,\hat b$ basis. While an exact description of the initial state and its evolution involves the detailed form of the Bogolubov coefficients, one can see the basic structure  from the preceding discussion.
As $t$ increases, the paired excitations of \eqref{pairstate} (or its wavepacket version) in the near horizon region evolve into the far regions, while maintaining their correlation and thermal spectrum.  And, once a localized excitation reaches $|r-R|\gg 1$, due to the asymptotic linear relation \eqref{Asympy} between $y^-$ or $\hat y^-$ and $r$, it is to a good approximation equally well-described as a correspondingly localized $a_k^\dagger$ excitation on the vacuum $|0\rangle_r$.  

So, in summary, we can describe the long-time evolution by converting an initial vacuum $|0\rangle_r$  into the energy eigenbasis, using the simplicity of the latter to evolve for finite time, and then converting back into the regular basis, with mode functions $e^{ikr}$.   The resulting state has a thermal spectrum for the excitations, and also has the entanglement structure between internal and external excitations exihibited in \eqref{pairstate}.  Tracing over internal excitations then explicitly leads to an external density matrix\cite{GiNe}, which has  a von Neumann entropy that grows in the expected linear fashion with time, as Hawking quanta are emitted.

The Hawking radiation -- and corresponding excitations inside the BH -- also make a contribution to the stress tensor; for a given state $|\psi\rangle$ of the Hawking radiation, we may for example calculate $\langle \psi|T_{--}|\psi\rangle$.  If we are interested in the asymptotic flux, then $T_{\mu\nu}$ is normal ordered with respect to the asymptotic coordinate $y_-$.  For example, with the initial vacuum state $|0\rangle_{-}$, the expectation value of the stress tensor may be found by comparing to $T_{\mu\nu}$ normal ordered with respect to the $x^-$ coordinates, as shown explicitly in \cite{SGTrieste}.  This gives
\beq\label{Tvev}
{}_{-}\langle0|:T_{--}:_y|0\rangle_{-} = \frac{1}{48} \left[1-\frac{1}{(1+\Delta e^{y^-})^2}\right]\ .
\eeq
One may alternately arrive at this result by using the formula for the conformal anomaly\cite{ChFu}\cite{CGHS} (for an overview, see \cite{SGBoltz}).
The expression \eqref{Tvev} exhibits evolution from the transitional, time-dependent, regime, to the stationary regime with constant flux.  Of course, the flux also implies that the BH shrinks, altering the time-dependent background geometry.  However, this effect, studied for example in \cite{CGHS},\footnote{For a review with further references, see {\it e.g.} \cite{Astrorev}.} is small over moderate times for large BHs.  Specifically, the backreaction of the flux on the geometry is suppressed in the small parameter $e^{2\phi}$ evaluated at the horizon, as can be seen from the equations of motion \eqref{meteqn}.

\subsubsection{Evolution and ``freezing" on nice slices}\label{freezesec}

While the straight slices, with $S(r)=r$, simplify aspects of the description, they do not give a complete description of the Hilbert space in four and higher dimensions, as is illustrated in Fig.~\ref{Fig4d}, since they intersect $r=0$.   In order to maintain a complete description of the state, one needs the slices to avoid $r=0$; sufficiently smooth such slices have been called nice slices\cite{Waldnice,LPSTU}.  So, we also need to understand evolution on this kind of slicing.  In our description \eqref{slicedef} of stationary slices, this behavior is achieved by choosing a slice function $S(r)$ that asymptotes to $-\infty$ at some finite $r=R_n$.  

Given such a slicing, one can once again consider different pictures based on different coordinates along the slices.  The coordinate $r$ can be used, but gives a nearly degenerate description near $r=R_n$.  This behavior may be avoided by instead using the coordinate $\rho=S(r)$.  

The accumulation of slices at $r=R_n$, illustrated in Fig.~\ref{Fig4d}, suggests that in a nice-slice description the wavefunction describing infalling excitations freezes at large times\cite{QBHB,BHQIUE}.  However, this does depend on the picture that is chosen.  For a picture based on the coordinates $r$ or $\rho=S(r)$, an infalling excitation $\exp\{-i\omega x^+\}$ has nonvanishing $t$ derivative at constant $r$ or $\rho$, and correspondingly the hamiltonian $H_\xi$ doesn't vanish; this is of course implied by conservation of the hamiltonian, and is also true for right-movers.  One way to understand this is from behavior of the lapse and shift: while, from \eqref{ADMvar}, $N\rightarrow0$ at $r=R_n$, the norm squared of the shift doesn't vanish.
This encodes the fact that while the slices accumulate, in these coordinates evolution shifts an excitation along the slices.

The freezing behavior can however be illustrated in a picture based on a different choice of coordinate.  A class of such coordinates along the slices takes the form\footnote{I thank J. Perkins for discussions on this class of coordinates.}
\beq\label{newcoord}
x=x^+ + g(r)\ ,
\eeq
with a general function $g(r)$.  In this case, the time derivative of an ingoing solution  $\exp\{-i\omega x^+\}$ at constant $x$ vanishes as $r\rightarrow R_n$.
 This is one example of the freezing behavior, which follows more generally from the form of the lapse and shift.  With the coordinate \eqref{newcoord} these become
\beq
N^2=\frac{1}{S'(2-fS')}\quad ,\quad N^x=\frac{g'-S'-fS'g'}{S'(2-fS')}
\eeq
and so vanish as the slices accumulate at $r=R_n$, where $S'\rightarrow\infty$.  

As a result, in such a picture the part of the wavefunction that falls into the BH ultimately does become time independent, or freeze, as can be seen for example from asymptotic vanishing of the hamiltonian \eqref{Fham}.  This type of picture therefore has the additional advantage of giving a particularly simple way to keep track of the excitations (and their information) inside of the BH -- they become static.
However, the choice of coordinates \eqref{newcoord} does not match the standard $t,r$ coordinates at large $r$, leading to a more complicated description there, unless one instead uses a time-dependent function $g(r,t)$, which then produces more complicated behavior.  Further treatment of evolution on nice slices will be left for future work.

\section{Evolution of internal excitations}

A slice-based Schr\"odinger description also offers a way to  understand better the evolution of excitations inside a BH.  This is true not just for infalling excitations, but also for the internal partners of the Hawking quanta, about which various past questions have been raised.

Here we will focus on characterizing some aspects of the latter ``Hawking partners,"  and their relation to the different quantum states of the BH.  We expect to be able to do so, since given a slicing and a choice of spatial coordinates, we have found a complete description of the evolution of the state (at leading perturbative level) through a hamiltonian \eqref{Fham} or \eqref{HT}.  

\subsection{Description of the state}

Consider the state at a long time after BH formation; again refer to Fig.~\ref{Figregions}, where we focus on the nearer region which the transitional radiation has left.  
In this region, as discussed above, the state is well-approximated as $|0\rangle_{X^-}$.  To describe its local properties,  one can work with localized wavepackets formed by superposing the modes $e^{ikr}$, or with the energy eigenmodes \eqref{Emodes}.  It is convenient to describe the resulting state, for example, as built on the regular vacuum $|0\rangle_r$.  
As noted above, far from the horizon, $|r-R|\gg 1$, the asymptotic linear coordinate relations \eqref{Asympy} imply that 
 the energy eigenmodes are to a good approximation identified with these regular modes.
This means that for purposes of studying excitations in these regions, the vacua $|0\rangle_r$ and $|\hat 0,0\rangle$ have nearly the same behavior, and the excitations $a^\dagger_k$ are approximately identified with $b^\dagger_k$ or $\hat b^\dagger_k$, depending on whether one is outside or inside the horizon.\footnote{Here, again, to make these statements precise, one should consider wavepacket superpositions of these operators in order to spatially localize.}  

This  also means that, well outside the BH atmosphere region at $|r-R|\lesssim 1$, the state has the same form as \eqref{pairstate},
\beq\label{LTstate}
|\Psi,T\rangle\simeq |0\rangle_{X^-} \sim C \sum_{\{n_k\}} e^{-\pi\int dk k n_k} |\{\widehat{n_k}\},\{n_k\}\rangle\ ,
\eeq
where the states $|\{\widehat{n_k}\},\{n_k\}\rangle$ are built on the regular vacuum $|0\rangle_r$ by acting 
with the operators $a^\dagger_k$ to create excitations inside or outside the horizon, and where strictly speaking one should therefore use wavepackets to localize.  So, in short, we see the Hawking excitations, with correlated partners created by (wavepacket superpositions of) $a^\dagger_k$ acting on $|0\rangle_r$, inside the horizon.

One way to better understand the nature of this state and these partner excitations is to describe their appearance to an infalling observer.  

\subsection{Infalling observers}

Bearing in mind the ultimate goal to better understand the $D$-dimensional case, we can investigate aspects of the state and excitations as seen by an inertially infalling observer moving in two dimensions, following a timelike geodesic.  The static BH metric 
\beq
ds^2=-f dx^{+2} + 2 dx^+ dr
\eeq
has  Killing vector $\xi^\mu=(1,0)$, implying conservation of $e=-\xi\cdot u$, with velocity $u^\mu = dx^\mu/d\tau$. For an observer starting at rest at infinity, this results in the equation of motion
\beq
\left(\frac{dr}{d\tau}\right)^2=1-f(r)\ 
\eeq
which is valid for radial infall in the general $D$-dimensional case.   In 2d this gives
\beq
u^r=\frac{dr}{d\tau}=-e^{R-r}
\eeq
with solution
\beq
r-R= \ln(\tau_a-\tau)\ .
\eeq
The observer reaches $r=-\infty$ at finite proper time $\tau=\tau_a$.  
The component $u^+$ is easily found,
\beq
u^+=\frac{\tau_a-\tau}{1+\tau_a-\tau} = \frac{e^{r-R}}{1+e^{r-R}}\ ,
\eeq
and integrates to give trajectory
\beq
x^+=\tau-\tau_a + \ln(1+\tau_a-\tau) + x_a^+ = -e^{r-R} +\ln\left(1+e^{r-R}\right) + x_a^+\ ,
\eeq
asymptoting to finite $x^+=x_a^+$ as $r\rightarrow-\infty$.  The diverging radial velocity $u^r$ at $r=-\infty$ occurs instead at $r=0$ in the higher-dimensional case, but the present study is a useful warmup exercise for that case.  It will also be useful to have the trajectory in the $\hat y^-$ and $X^-$ coordinates, related through \eqref{basicc}, \eqref{yhatdef}, and \eqref{Krusk},
\beq
\hat y^- = \tau_a-\tau -x_a^++2R + \ln(1-\tau_a+\tau)\ ,
\eeq
\beq\label{Xmtraj}
X^-=(1-\tau_a+\tau) e^{2R-x_a^+-(\tau-\tau_a)}\ ,
\eeq
and giving velocity components in $(x^+, \hat y^-)$ coordinates
\beq
u^\mu=\left(\frac{1}{e^{R-r}+1},\frac{1}{e^{R-r}-1}\right)\ .
\eeq

\subsection{Infalling observations}

From the preceding section, recall that the state is well-approximated as the Unruh state $|0\rangle_{X^-}$ in the stationary region.  The relation \eqref{Xmtraj} shows that definite frequency $X^-$ modes are not definite frequency in the observer's proper time $\tau$, and in the strong-coupling limit $\tau\rightarrow \tau_a$, $r\rightarrow-\infty$, 
\beq
\frac{dX^-}{d\tau}\rightarrow 0\quad ,\quad \frac{d^2X^-}{d\tau^2}\rightarrow -e^{2R-x_a^+}\ .
\eeq
This implies that the Unruh vacuum $|0\rangle_{X^-}$ appears excited to the infalling observer.  Likewise, in this limit
\beq
\frac{d\hat y^-}{d\tau}\rightarrow 0\quad ,\quad \frac{d^2\hat y^-}{d\tau^2}\rightarrow -1\ ,
\eeq
and so the vacuum $|0\rangle_r$, which looks like $|\hat 0,0\rangle$ for $|r-R|\gg 1$, also appears excited to the infalling observer.

From the form of the state \eqref{LTstate} we see that the physical state $\simeq|0\rangle_{X^-}$ has $a^\dagger_k \simeq \hat b_k^\dagger$ excitations above $|0\rangle_r$ in the internal region $R-r\gg 1$.  To characterize their appearance to the infalling observer, we can find their energy and momentum seen by such an observer, by taking inner products with the dyad carried by that observer.  The observed energy of one of these $a^\dagger_k \simeq \hat b_k^\dagger$ excitations is given by 
\beq
\omega_{obs}=-k_\mu u^\mu =  k\frac{1}{e^{R-r}-1}\ ,
\eeq
and so is positive and vanishes as the observer asymptotes to strong coupling.  The observed momentum, found from taking the product with the unit spatial vector satisfying  $e_\mu u^\mu=0$, which has  $x^+,\hat y^-$ components
\beq
e_\mu=\hf(1+e^{R-r},1-e^{R-r})=\frac{1}{2}(1+\frac{1}{\tau_a-\tau}, 1-\frac{1}{\tau_a-\tau})\ ,
\eeq
 is likewise $k_{obs}= e_\mu k^\mu = \omega_{obs}$.

\subsection{Properties and creation of internal excitations}

While the internal $a^\dagger_k \simeq \hat b_k^\dagger$ excitations have positive energy as seen by a local infalling observer, we saw they have negative energy for the global hamiltonian \eqref{Hb} governing  evolution along the Killing vector $\xi$.  This contrast adds emphasis to the role of different hamiltonians, corresponding to different pictures.  Their negative global energy is important in the production of the Hawking radiation, but does raise some potential puzzles regarding the nature of the states of the BH.

In particular, the existence of negative energy excitations suggests the possibility of an unphysical spectrum with an unbounded number of BH states of finite energy, which would certainly exceed a number of states given by exponentiating the Bekenstein-Hawking entropy.  Specifically, given a BH state $|\psi\rangle$, consider
\beq\label{degstate}
|\psi\rangle \rightarrow \prod_{ij} \tilde a_i^\dagger \hat b_j^\dagger |\psi\rangle\ ,
\eeq
where $\tilde a_i^\dagger$ and $\hat b_j^\dagger$ are internal wavepacket superpositions of the operators $\tilde a^\dagger_k$ and $\hat b_k^\dagger$, respectively.  The form of the hamiltonian \eqref{LHam} and \eqref{Hb} suggests that it is possible to construct an infinite number of states of this kind that are degenerate with $|\psi\rangle$ in the definite energy/plane wave limit, by matching energies of the excitations $\tilde a_k^\dagger$ and $\hat b_k^\dagger$.
This, in turn, could result in unphysical phenomena, such as infinite pair production of BHs\cite{WABHIP}.  Moreover, this suggests that there is a large number of low-energy states that are near-horizon, short wavelength states\cite{AMPSS}; if it is assumed that these are excited by the mechanism that corrects the Hawking process to restore unitarity, that leads to singular behavior near the horizon\cite{SGTrieste,Brau,AMPS},  called ``firewalls" in \cite{AMPS}.

Key questions, however, are how such states might be produced, and whether they are truly physically realizable states.

We have seen clearly, {\it e.g.} from  the hamiltonian \eqref{Hm} or state \eqref{LTstate}, that internal $a_k^\dagger\simeq \hat b^\dagger_k$ excitations are produced in the Hawking process, but always in correlated pairs with the outgoing Hawking particles.  Global energy conservation is satisfied since the outgoing particle and internal excitation are positive and negative energy, respectively.  The internal excitation can be interpreted as lowering the energy of the BH.

This then raises the question of the role of such BH states {\it without} their correlated outside partners.  For example, consider the state that, at a given time, is described as an internal wavepacket superposition of $a_k^\dagger\simeq \hat b^\dagger_k$ acting on $|0\rangle_{X^-}$.  At earlier times, this corresponds to a state with arbitrarily high wavenumbers $k$, very close to the horizon, as can be seen for example by considering the evolution of the mode $\phi_k=e^{ik(r-R)}$ backward in time from $T_0$; in the straight slicing one finds\cite{SEHS}
\beq
\phi_k(T,r) = e^{ik \sinh^{-1} \left[ e^{T_0-T}\sinh(r-R)\right]}\ .
\eeq
Only if specific collections of such modes, like in \eqref{LTstate}, merge with a corresponding collection outside the horizon does one find a 
regular state in the far past; in this case, one can alternately understand this by examining cancellations in the stress tensor, and for example test these cancellations in gravitational scattering of an infalling object\cite{BHIUN}.  This is what happens with the evolution of the Hawking state and related non-singular states, as has been even more explicitly demonstrated by the evolution described in this paper: outside Hawking excitations and internal partners merge into a regular state when the evolution is run backwards.
But in a different state,
without such pairing, the state becomes ultraplanckian, and so does not clearly correspond to a physical state.

This suggests invoking a general principle that ``history matters" in determining whether or not a state is physical; a physical state must be producible by a physical process.  If so, for example, a state that when evolved backward yields an unphysical singularity, cannot be a physical state.  Put differently, hamiltonian evolution is not obligated to produce all states.  (Similar proposals were made in \cite{GiMa} and \cite{Pageext}.)

Of course, there are alternate ways to produce $\hat b^\dagger_k$ excitations, such as through emission by an infalling observer.  For a simple model, one can describe decay of an infalling mass $m$ particle into right and left moving massless excitations at $r=r_0$ inside the horizon.  In the local inertial frame of the infalling mass, the decay produces excitations with momenta
\beq
p_1^\mu=(m/2,m/2)\quad ,\quad p_2^\mu =(m/2,-m/2)\ ,
\eeq
which, in $(x^+,\hat y^-)$ coordinates correspond to
\beq
p_1^\mu= \left(\frac{m}{e^{R-r_0}+1},0\right) \quad,\quad p_2^\mu=\left(0,\frac{m}{e^{R-r_0}-1}\right) \ .
\eeq
respectively.  The first of these is a $\hat b_k^\dagger$ excitation with $k=m(e^{R-r_0}-1)/2$, with correspondingly negative global energy $-\xi\cdot p_1$.  The global energies are
\beq
E_0= m\quad,\quad E_1= \frac{m}{2}\left(1-e^{R-r_0}\right)\quad,\quad E_2=\frac{m}{2}\left(1+e^{R-r_0}\right)\ ,
\eeq
and are conserved, as expected.

Thus while such decay or emission processes excite the $\hat b^\dagger_k$ modes, they do not provide a new way to construct an arbitrarily large number of BH states with finite energy; the global BH energy increases by the energy of whatever infalling matter produces the excitation, and energy conservation implies correspondingly higher energy $\tilde a^\dagger_k$ excitations are produced inside the BH.

To summarize the discussion so far, in such internal emission processes, $\hat b_k^\dagger$ excitations are paired with {\it higher} energy $\tilde a_k^\dagger$ excitations, so do not necessarily lead to the large degeneracy of \eqref{degstate}, and the Hawking process only leads to production of $\hat b_k^\dagger$ excitations  paired with external $b_k^\dagger$ excitations, and not independently to states \eqref{degstate}.  This discussion is also expected to extend to higher dimensions.

One might consider other processes, for example thermal excitation, where $\hat b_k^\dagger$ excitations are produced entangled with a thermal bath.  However, such a thermal state must ultimately arise via evolution through underlying microphysical processes, such as we have described.  For example, a state like \eqref{degstate} can be created by a process in which one particle falls in, one particle is Hawking emitted, and then the process repeats.  This sequence only builds up $\sim \exp\{S_{BH}\}$ states after a very long time $t\propto S_{BH}$.  Another alternative, in the higher-dimensional context, is to produce states such as \eqref{degstate} through BH pair production\cite{Gibb,GaSt,DGKT,DGGH}.  

In either of the later cases, one also expects that we need to go beyond the approximation we have considered of neglecting the gravitational backreaction, and possibly other effects.  Once the emitted quanta are an appreciable fraction of the BH mass, and certainly by the time the BH has emitted $\calo(S_{BH})$ quanta, we expect to need to include such effects, whose leading contribution is the perturbative gravitational dressing.  And, once these effects are considered, that raises the possibility that they eliminate the na\"\i vely infinite degeneracy of states \eqref{degstate} that we have found by neglecting them.  In short, these states may not even exist to be consistently excited through thermal or pair production processes, or, there may not be a consistent physical channel to produce them.  At the least, a full treatment of these would seem to require including effects that go beyond the leading perturbative description of \eqref{degstate}, and in particular the full non-perturbative quantum theory may place additional limits on what states exist in the true Hilbert space.\footnote{One might also consider the role of $\hat b^\dagger_k$ excitations in loops\cite{AMPSS}, but ultimately this is also a question of which states consistently exist in the theory.}

One can of course describe such states at the perturbative level in a nice-slice description.  The excitations, {\it e.g.} due to successive Hawking emission, can be thought of as distributed sequentially along the slice.  Moreover, in a picture like that described in \ref{freezesec}, the internal part of the state can be taken to be time-independent, or freeze.  However, once again, we expect a complete treatment of these states requires a full accounting of gravitational backreaction, as well as other quantum-gravitational effects that go beyond this perturbative description.

\section{Conclusion and directions}

This paper has described the evolving quantum state of a black hole, in the dilaton gravity model of \cite{CGHS}, beginning with the full gravitational theory and then considering the leading perturbative approximation neglecting gravitational backreaction.  We expect many aspects of this treatment to apply to higher-dimensional black holes, with modifications that are relatively minor\cite{GiPe}, at least in the perturbative approximation.  

The present treatment has the virtue of being extendable also to the case of interacting field theories, at least in principle.  Whereas Hawking's original treatment\cite{Hawk} involved tracing back a particular mode to near the horizon, by following its free propagation, the slice-based evolving state that we have described can in principle incorporate interactions.  Specifically, there could also be interaction terms in the expressions for the hamiltonian \eqref{Fham} or \eqref{HT}, which would be incorporated into the description of the evolving state.  That would of course result in more complicated expressions than the quadratic hamiltonian \eqref{Hm}, with higher-order terms encapsulating the interactions.

In the present analysis, in particular we have seen explicitly both the leading perturbative treatment of the internal states of the BH, and their excitation in a state entangled with the Hawking modes in the process of the time evolution via the Hawking effect. 

This of course only makes more concrete the basic conundrum of Hawking radiation.  One builds up large entanglement between the internal excitations of the BH and the Hawking radiation.  If this picture persists once gravitational backreaction is included, and the BH disappears at the end of evolution, that is inconsistent with unitarity.

A central question, then, is how Hawking evolution is modified.  The treatment of this paper leads to a more complete foundation on which to base description of effects that go beyond this evolution, to give evolution consistent with unitarity.  Key questions are when we cease to believe the time evolution that we have described, and due to what new effects.  For example, with such an explicit description of the ``standard" quantum field theory evolution, one can ask how precisely new proposals, such as those of 't Hooft\cite{tHooft}, of replica wormholes\cite{SSS,PSSY,AHMST,MaMa}, or those of nonviolent unitarization\cite{NVU,BHQU} alter that evolution.  A traditional field-based quantization apparently cannot unitarize the evolution, so one expects these new effects to be truly quantum gravitational in nature.

Taking a step back, one can characterize the problem in a more general fashion.  Specifically, if a BH can be described as a quantum subsystem, and builds up a large amount of entanglement with the complementary environment subsystem, and then disappears, that is inconsistent with unitarity.  The leading perturbative description given in this paper exhibits BHs as such quantum subsystems, described in the usual fashion for QFT.  Specifically, beginning with a chosen BH vacuum state $|0\rangle$, one can consider independent excitations of internal and external modes, {\it e.g.}
\beq
\prod_{ij} \hat b_i^\dagger b_j^\dagger|0\rangle
\eeq
associated for example to wavepackets with support purely inside or outside the horizon, and likewise for left-moving excitations built from $\tilde a_k^\dagger$.  The hamiltonian, \eqref{Fham} or \eqref{HT}, can likewise decomposed into a piece with support inside the BH, a piece outside, and a piece supported at the horizon that provides an interaction between the two.  

Of course, inclusion of even the leading gravitational backreaction modifies this structure\cite{DoGi2,DoGi3} (see also \cite{CGPR}), since the gravitational dressing associated to an internal excitation extends outside the horizon\cite{GiWe}.  However, the arguments of \cite{Gisub} suggest that there is still a subsystem structure, since the gravitational degrees of freedom may be taken to be in a state where the external structure of the dressing depends only the total Poincar\'e charges of the internal matter, and not on other features of the internal state.  (This could also be more concretely explored in an extension to this paper.)

Then, if this subsystem structure persists in the perturbative theory, along with the entanglement built up by the Hawking effect, the contradiction with unitarity remains.  New quantum-gravitational effects, outside of this perturbative description, are apparently needed to restore unitary evolution.  These effects need to either modify this subsystem structure, or to transfer information (more specifically, entanglement) from the BH subsystem to the environment.

A key question for any proposed unitarization of BH evolution is thus to describe these new effects, as a departure from the standard perturbative evolution like that described in this paper.  For example, this test could be applied to the proposal of 't Hooft\cite{tHooft}, to ask how his proposed evolution differs from the standard LQFT evolution described above.\footnote{In particular, while some of the discussion of \cite{tHooft} takes more of a particle focus, the present treatment emphasizes the role of a QFT-based field focus on the evolution of the quantum state.}  Alternately, a newer development is the proposal that replica wormhole effects lead to an ultimately unitary description, as suggested by an argument that a calculated entropy falls to zero at the end of evolution\cite{SSS,PSSY,AHMST,MaMa}.  An important question is to understand this claim at the level of quantum amplitudes, and their modification to the standard evolution.

More specifically, the LQFT evolution with hamiltonian \eqref{Fham} or \eqref{HT} leads to buildup of entanglement between the BH and environment, and allows transfer of information from the environment to the BH, but due to the underlying local/causal nature of its evolution does not transfer entanglement from the BH to the environment.  If the BH is thought of, in an appropriate approximation, as a subsystem, unitarity {\it requires} modifications to this hamiltonian that can transfer entanglement from BH to environment.  One can ask how these modifications are described, in a given proposal.

We might anticipate that such new quantum gravity effects are ``small," and if they are a small departure from the conventional LQFT description, we expect to be able to parameterize them as a deviation from that description.
This is the basic approach of \cite{SGmodels}\cite{BHQIUE}\cite{NVNL,NVUEFT,NVNLT}\cite{NVU}\cite{BHQU}.  Specifically, if there must be additional interactions between the BH quantum state and the environment of the BH, one can parameterize them in terms of an additional contribution to the hamiltonian that transfers entanglement.  The simplest such structure that does so bilinearly couples operators that act on the BH state to operators that act on the environment.  The natural scale for these operators to act is that of the horizon radius, $\calo(R)$, rather than on a highly tuned scale $R+\calo(l_{\rm Planck})$ or on a much longer scale $\gg R$.\footnote{In fact the ``firewall" picture advocated in \cite{AMPS} implicitly assumes that Hawking evolution is modified only once excitations are traced back into a near-horizon, planckian regime.  We have seen that for Hawking evolution the role of  such a regime is  basis dependent and not necessarily fundamental. This suggests that unitarizing corrections to Hawking evolution quite plausibly operate on longer scales.} These statements, plus a motivated assumption of universality of these couplings, greatly simplify the problem of parameterizing these interactions, and one can investigate the extent to which the necessary entanglement transfer can be accomplished by small interactions\cite{NVU}\cite{GiRo}\cite{BHQU}.  

This approach may in fact give an effective description of different underlying mechanisms for unitarization.  For example, if the mechanism is related to baby universe emission, as argued in connection with work on replica wormholes by \cite{MaMa}, then a possible structure for corrections to the hamiltonian is of the form
\beq
\Delta H = A_i^\dagger \calo_i + {\rm h.c.}\ ,
\eeq
where $\calo_i$ is the operator describing the effect of baby universe emission and $A_i^\dagger$ creates a baby universe, as in \cite{Cole,GiSt}. 
If this is an important effect for baby universes of size comparable to the BH scale $R$, then the operators 
$\calo_i$ are expected to be nonlocal on this scale.  Then, for example, in a baby universe state that diagonalizes the creation and annihilation operators, as in \cite{GiSt,Cole}, one has induced couplings to such nonlocal operators present in the evolution.

Or, possibly the information transferring interactions are due to other quantum-gravitational effects modifying spacetime structure, {\it e.g.} as in \cite{QFG}.  Whatever the origin of such quantum gravity interactions, the  description of the (approximate) LQFT evolution given in this paper provides a concrete foundation for parameterization of such departures from that evolution, capable of the necessary transfer of entanglement and restoration of unitarity.

\vskip.3in
\noindent{\bf Acknowledgements} 
 
This material is based upon work supported in part by the U.S. Department of Energy, Office of Science, under Award Number {DE-SC}0011702.  I thank T. Jacobson and J. Perkins for useful conversations.

\appendix

\section{Useful ADM expressions}
\label{ADMA}

This appendix will collect some useful formulas for the ADM approach to gravity, in $D$ dimensions.  The ADM form of the metric is 
\beq
ds^2 = -N^2 dt^2 + q_{ij} (dx^i + N^i dt)(dx^j + N^j dt)\ , 
\eeq
 with lapse $N$, shift $N^i$, and spatial metric $q_{ij}$, and the inverse metric is
\beq
g^{\mu\nu} = 
\begin{pmatrix}
-{1/N^2} & N^i/N^2\\
N^i/N^2& q^{ij} - {N^i N^j}/{N^2}\\
\end{pmatrix}\ .
\eeq
With the definition  $q={\rm det}\,q_{ij}$,
the volume element is
\beq
\sqrt{|g|} =N\sqrt q\ .
\eeq
The unit normal to the constant-$t$ slices is
\beq
n^\mu = (1,-N^i)/N\ .
\eeq

To find the gravitational action, we need the Ricci scalar, which takes the form \cite{KucharR,KRV}
\beq
\sqrt{|g|} R = N\sqrt{q} \left[K_{ij}K^{ij} -K^2 +R(q)\right] - 2\sqrt{q} D^2 N + 2\partial_i\left(\sqrt{q} K N^i \right) - 2\partial_t\left(\sqrt{q} K\right)\ .
\eeq
Here  spatial indices are raised and lowered with $q_{ij}$,  $D_i$ is the spatial covariant derivative computed using $q_{ij}$, and
the extrinsic curvature of the constant-$t$ slices is, with dot denoting $\partial_t$,
\beq
K_{ij} = -\frac{1}{2N} \left(\dot q_{ij} - D_i N_j - D_j N_i\right)\ .
\eeq
The gravitational action resulting from these expressions is investigated in the main text.

We also need to express scalar field actions in ADM variables, which is done via
\beq
(\nabla \phi)^2 = -\frac{1}{N^2}\left(\dot \phi - N^i\partial_i\phi\right)^2 + q^{ij}\partial_i\phi\partial_j\phi = -(\partial_n\phi)^2+ q^{ij}\partial_i\phi\partial_j\phi\ 
\eeq
which we have rewritten in terms of the normal derivative,
\beq
\partial_n\phi=n^\mu\partial_\mu\phi  =\frac{\partial_t \phi - N^i\partial_i \phi}{N}\quad \Leftrightarrow \quad \partial_t\phi = N\partial_n\phi + N^i \partial_i \phi \ .
\eeq
This gives a massless scalar action
\beq
S_\phi=-\hf \int d^D x \sqrt{|g|} (\nabla\phi)^2 =  \frac{1}{2} \int dt d^{D-1}x \sqrt q N \left[ (\partial_n\phi)^2 - q^{ij}\partial_i \phi \partial_j \phi\right]\ ,
\eeq
where here we use the standard normalization (a different normalization is used in the main text).

The canonical momentum is 
\beq
\pi_\phi= \frac{1}{\sqrt q}\frac{\delta S_\phi}{ \delta \dot\phi} = \frac{1}{N}\left(\dot \phi - N^i\partial_i\phi\right) = \partial_n\phi\ ,
\eeq
from which we find the canonical form of the action,
\beq
S_\phi=\int dt d^{D-1}x \sqrt q \left(\pi_\phi \dot\phi  - \calh_\phi\right)\ ,
\eeq
with hamiltonian density
\beq\label{canham}
\calh_\phi= \frac{N}{2} \left( \pi_\phi^2+q^{ij}\partial_i\phi \partial_j\phi\right)+ \pi_\phi N^i \partial_i\phi\ .
\eeq
The hamiltonian is then
\beq
H=\int d^{D-1}x \sqrt{q}\left[ \frac{N}{2}\left(\pi_\phi^2+q^{ij}\partial_i\phi \partial_j\phi\right) + \pi_\phi N^i \partial_i\phi \right]
=
\frac{1}{2} \int d^{D-1}x \sqrt{q} N\left[ \frac{\dot\phi^2}{N^2}  + g^{ij}\partial_i\phi\partial_j \phi\right]\ .
\eeq

\section{Time dependent backgrounds and perturbation picture}
\label{PP}

This appendix provides a simple description of quantum evolution that is a perturbation of a dynamic background, and of the resulting transformation to what can be called ``perturbation picture."

For a simple example to illustrate the basic idea of perturbation picture, consider the hamiltonian of a particle in a potential,
\beq
H=\frac{p^2}{2}+ V(q)\ ,
\eeq
and suppose that the equations of motion
\beq\label{EOM}
p=\dot q\quad ,\quad \dot p = -V'(q)
\eeq
have a dynamic solution $p_0(t)$, $q_0(t)$.  We would like to consider quantization of fluctuations that are perturbations of this classical solution,
\beq
p= p_0 +\tilde p\quad ,\quad q=q_0+\tilde q\ .
\eeq
The hamiltonian can be expanded as 
\beq
H=H_0 + \dot q_0 \tilde p - \dot p_0 \tilde q + \tilde H(\tilde p,\tilde q) 
\eeq
where we use the equations of motion \eqref{EOM} and where
\beq
 \tilde H(\tilde p,\tilde q)  = \frac{\tilde p^2}{2} +  \sum_{n=2}^\infty \frac{1}{n!} \tilde q^n V^{(n)}(q_0)\ .
 \eeq
 
In the standard Heisenberg picture, the equations of motion \eqref{EOM} become the Heisenberg equations,
\bea
\partial_t q_H &=& i [H,q_H] = i [H,\tilde q_H] = \dot q_0 + i [\tilde H,\tilde q_H]\quad \Rightarrow\quad  \partial_t\tilde q_H=i[\tilde H,\tilde q_H]\ ,\\
\partial_t p_H &=& i [H,p_H] = i [H,\tilde p_H] = \dot p_0 + i [\tilde H,\tilde p_H]\quad \Rightarrow\quad  \partial_t\tilde p_H=i[\tilde H,\tilde p_H]\ ,
\eea
where the subscript ``H" denotes the Heisenberg picture.  In the Heisenberg picture, the state is time-independent, $|\psi(t)\rangle_H=|\psi(0)\rangle_H$
and so the expectation value of a general operator $\tilde A(\tilde p, \tilde q)$  evolves as
\beq\label{MEevol}
\langle\psi|\tilde A|\psi\rangle = \langle \psi(0)| \tilde A_H(t) |\psi(0)\rangle =  \langle \psi(0)| \tilde U^{-1}(t) \tilde A_H(0) \tilde U(t) |\psi(0)\rangle 
\eeq
where $\tilde U(t)$ is the unitary evolution operator satisfying
\beq
\partial_t \tilde U(t)= i[\tilde H,\tilde U(t)]\ .
\eeq

Perturbation picture can now be defined analogously to the standard Schr\"odinger picture, as a picture in which the operator $\tilde A$ does not evolve but all evolution is in the state.  Specifically, define the perturbation picture state and operators as
\beq
|\tilde \psi(t)\rangle_P = \tilde U(t) |\psi(0)\rangle\quad ,\quad \tilde A_P= \tilde A_H(0) =  \tilde U(t) \tilde A_H(t) \tilde U^{-1}(t)\ ,
\eeq
which clearly reproduce the evolution \eqref{MEevol}.
These satisfy 
\beq
i\partial_t |\tilde \psi(t)\rangle_P =\tilde H  |\tilde \psi(t)\rangle_P\quad ,\quad \partial_t \tilde A_P=0\ 
\eeq
so evolution of the state is governed by $\tilde H$.  This picture can be contrasted with the standard Schr\"odinger picture, in which the state instead
evolves as
\beq
| \psi(t)\rangle_S = U(t) |\psi(0)\rangle\ ,
\eeq
with the evolution operator $U$ determined by the {\it full} hamiltonian $H$.  Perturbation and Schr\"odinger pictures are clearly identical for time-independent backgrounds.  
For time-dependent backgrounds, another way to see their distinction is to notice that in Schr\"odinger picture, time-independence of $q_S, p_S$ implies that
\beq
\partial_t \tilde q_S = -\dot q_0 \quad ,\quad \partial_t \tilde p_S = -\dot p_0\ ,
\eeq 
as contrasted with the time-independent operators of perturbation picture.  In the classical context, the difference can be understood in terms of a canonical transformation\cite{MaroUH}.

This general structure can clearly be continued to more complicated systems, {\it e.g.} involving evolution of quantum fields as perturbations of a time-dependent background, as in the main text.

\section{Bogolubov transformation -- straight slicing}
\label{BOG}

This appendix describes an example of the Bogolubov transformation that relates the energy eigenmodes to regular modes, in the special case of regular modes defined using the straight slicing, $S(r)=r$.  

On a fixed $t$ slice, with slicing given by $x^+=t+r$, the right-moving field operator can be expanded either in terms of the regular basis $e^{ikr}$ or the energy eigenbasis $e^{-i\omega y^-}$,  $e^{-i\omega \hat y^-}$:
\begin{align}
\label{Freln}
F&=\int_0^\infty \frac{dk}{{\sqrt {2\pi}}2 k} (a_k e^{ikr} + a_k^\dagger e^{-ikr} )\\
&= \theta(r-R) \int_0^\infty \frac{d\omega}{\sqrt{2\pi}2\omega} (b_\omega e^{-i\omega y^-} + b_\omega^\dagger e^{i\omega y^-} )+\theta(R-r) \int_0^\infty \frac{d\omega}{\sqrt{2\pi}2\omega} (\hat b_\omega e^{-i\omega \hat y^-} + \hat b_\omega^\dagger e^{i\omega \hat y^-} )  \ .\nonumber
\end{align}
The Bogolubov coefficients relating the modes can be defined by 
\bea
 \theta(r-R)e^{-i\omega y^-} &=& \int_0^\infty dk (\alpha_{\omega k}^+ e^{ikr} + \alpha^-_{\omega k} e^{-ikr})\cr
 \theta(R-r)e^{-i\omega \hat y^-}  &=& \int_0^\infty dk (\hat \alpha_{\omega k}^+ e^{ikr} + \hat \alpha_{\omega k}^- e^{-ikr})\ .
\eea
From these we find 
\beq
\alpha_{\omega k}^\pm = \frac{1}{2\pi}  \int_R^\infty dr e^{-i\omega y^- \mp ikr}= \frac{e^{-i\omega(t-R)\mp i k R}}{2\pi} \int_0^\infty dx e^{i\omega \ln(2\sinh x) \mp ikx}
\eeq
and
\beq
\hat \alpha_{\omega k}^\pm = \frac{1}{2\pi}  \int_{-\infty}^R dr e^{-i\omega \hat y^-\mp ikr} 
= \frac{e^{i\omega(t-R)\mp i k R}}{2\pi} \int_0^\infty dx e^{-i\omega \ln(2\sinh x) \pm ikx}\ ,
\eeq
where the coordinate relations  \eqref{LTcoord} and simple changes of variables  have been used.  These can be evaluted to give
\beq
\alpha^\pm_{\omega k} =\frac{e^{-i\omega(t-R)\mp i k R}}{4\pi} \frac{\Gamma(1+i\omega)\Gamma\left[-i(\omega\mp k)/2 + \epsilon\right]}{\Gamma\left[1+i(\omega\pm k)/2\right]} 
\eeq
and 
\beq
\hat \alpha^\pm_{\omega k} =\frac{e^{i\omega(t-R)\mp i k R}}{4\pi} \frac{\Gamma(1-i\omega)\Gamma\left[i(\omega\mp k)/2 + \epsilon\right]}{\Gamma\left[1-i(\omega\pm k)/2\right]} 
\eeq
where $\epsilon>0$ is a small convergence factor (which gives the pole prescription); also note that these may be rewritten in terms of a beta function.
This leads to a relation 
\beq
a_k = \int_0^\infty d\omega\frac{k}{\omega} \left( b_\omega \alpha_{\omega k}^+ + b_\omega^\dagger \alpha^{-*}_{\omega k} + \hat b_\omega \hat \alpha^+_{\omega k} + \hat b_\omega^\dagger \hat \alpha^{-*}_{\omega k} \right)\ .
\eeq
and likewise for creation operators, between the ladder operators.

\mciteSetMidEndSepPunct{}{\ifmciteBstWouldAddEndPunct.\else\fi}{\relax}
\bibliographystyle{utphys}
\bibliography{sch}{}

\end{document}